\documentclass{sn-jnl}

\usepackage{graphicx}%
\usepackage{multirow}%
\usepackage{amsmath,amssymb,amsfonts}%
\usepackage{amsthm}%
\usepackage{mathrsfs}%
\usepackage[title]{appendix}%
\usepackage{xcolor}%
\usepackage{textcomp}%
\usepackage{manyfoot}%
\usepackage{booktabs}%
\usepackage{algorithm}%
\usepackage{algorithmicx}%
\usepackage{algpseudocode}%
\usepackage{listings}%

\usepackage{lineno}
\usepackage{etoolbox,setspace}
\usepackage{bm}
\graphicspath{{Figures/}}
\usepackage[superscript,biblabel,nomove]{cite}
\usepackage{jabbrv}
\usepackage[section]{placeins} 
\usepackage{orcidlink}

\usepackage{subfiles}

\makeatletter
\renewcommand{\maketitle}{%
  \begingroup\singlespacing
    \@maketitle
  \endgroup
}
\makeatother

\makeatletter
\renewcommand{\addressfont}{%
  \reset@font\fontsize{10.5bp}{13.5bp}\selectfont\titraggedcenter%
}
\makeatother

\makeatletter
\@ifpackageloaded{geometry}{%
  \newgeometry{twoside=false, reversemp=false, bindingoffset=0mm, margin=25mm}%
}{}
\makeatother

\makeatletter
\g@addto@macro\abstractfont{\unboldmath}
\makeatother

\makeatletter
\renewcommand{\@tablecaption}[2]{%
  \begingroup
    \setlength{\@tempdima}{\dimexpr(\textwidth-\linewidth)/2\relax}%
    \noindent\hspace*{-\@tempdima}
    \parbox{\textwidth}{%
      \tablecaptionfont\raggedright
      {\bfseries #1}\hskip2mm #2\par
    }%
    \par
  \endgroup
}
\makeatother

\makeatletter
\renewcommand{\@biblabel}[1]{#1.}
\makeatother

\makeatletter
\def\email#1{%
  \global\advance\emailcnt by 1\relax
  \if@corauemail
    \ifx\corrauthemail\@empty \else \g@addto@macro\corrauthemail{; } \fi
    \g@addto@macro\corrauthemail{\textcolor{blue}{#1}}%
  \else
    \ifx\authemail\@empty \else \g@addto@macro\authemail{; } \fi
    \g@addto@macro\authemail{\textcolor{blue}{#1}}%
  \fi
}
\makeatother

\unnumbered

\begin{document}

\title[Article Title]{\textbf{Large language model-enabled automated data extraction for concrete materials informatics}}

\author[1,2,3]{\fnm{Zhanzhao} \sur{Li}~\orcidlink{0000-0001-7674-7424}}
\author[4]{\fnm{Kengran} \sur{Yang}~\orcidlink{0000-0001-9766-8063}}
\author[1,2,3]{\fnm{Qiyao} \sur{He}~\orcidlink{0009-0005-7496-9126}}
\author*[1,2,3]{\fnm{Kai} \sur{Gong}~\orcidlink{0000-0002-5884-4607}}
\email{\texttt{kg51@rice.edu}}

\affil[1]{\orgdiv{Department of Civil and Environmental Engineering}, \orgname{Rice University}, \orgaddress{\city{Houston}, \state{TX} \postcode{77005}, \country{USA}}}
\affil[2]{\orgdiv{Rice Advanced Materials Institute}, \orgname{Rice University}, \orgaddress{\city{Houston}, \state{TX} \postcode{77005}, \country{USA}}}
\affil[3]{\orgdiv{Ken Kennedy Institute}, \orgname{Rice University}, \orgaddress{\city{Houston}, \state{TX} \postcode{77005}, \country{USA}}}
\affil[4]{\orgname{Independent researcher}, \orgaddress{\city{Princeton}, \state{NJ}, \country{USA}}}

\abstract{The promise of data-driven materials discovery remains constrained by the scarcity of large, high-quality, and accessible experimental datasets. Here, we introduce a generalizable large language model (LLM)-powered pipeline for automated extraction and structuring of materials data from unstructured scientific literature, using concrete materials as a representative and particularly challenging example. The pipeline exhibits robust performance across a broad range of LLMs and achieves an $F_1$ score of up to 0.98 for diverse composition--process--property attributes. Within one hour, it extracts nearly 9,000 high-quality records with over 100 attributes from a corpus screened from more than 27,000 publications, enabling the construction of the largest open laboratory database for blended cement concrete. Machine learning analyses underscore the importance of large, diverse, and information-rich datasets for enhancing both in-distribution accuracy and out-of-distribution generalization to unseen materials. The proposed pipeline is readily adaptable to other materials domains and accelerates the development of scalable data infrastructures for materials informatics.}

\maketitle

\section{Introduction}

Data-driven approaches have become an increasingly important component of the materials science toolbox, enabling accelerated discovery, design, and optimization across broad classes of materials systems\cite{Agrawal2016,Morgan2020,Batra2020,Ramprasad2017}. A persistent bottleneck to fully exploiting their potential, however, is the scarcity of large, high-quality, and accessible datasets\cite{Olivetti2020,Schilling-Wilhelmi2024}. While substantial community efforts have led to a growing number of open materials databases, most of them are derived from computational methods\cite{Curtarolo2012,Saal2013,Jain2013,Draxl2019,Choudhary2020,Talirz2020,Batra2020}. Experimentally based databases remain comparatively rare and are typically limited in both size and scope\cite{Olivetti2020,Dagdelen2024}. Meanwhile, the scientific literature contains a wealth of peer-reviewed and reliable experimental data accumulated over decades of research, yet much of this information is not captured by existing databases. With the volume of publications expanding at an unprecedented pace, this vast body of literature represents a largely untapped resource for establishing large-scale, comprehensive data infrastructures to support materials informatics.

Despite this opportunity, extracting materials data at scale from the scientific literature remains challenging. Materials data are typically presented in highly unstructured and heterogeneous formats, dispersed across various sections of text and embedded within often complex tables and figures\cite{Olivetti2020,Kononova2021}. These challenges are further compounded by inconsistent units, non-standard naming conventions, and the widespread use of domain-specific acronyms, all of which require contextual interpretation and domain expertise to resolve. Consequently, manual data extraction is not only labor-intensive and time-consuming but fundamentally incompatible with the pace and scale of modern scientific publishing.

Text mining and natural language processing (NLP) techniques have been explored to automate the extraction of materials data from unstructured scientific literature \cite{Olivetti2020,Kononova2021,Swain2016,Shetty2022,Jensen2019,Kononova2019,Kim2017a,Wang2022a,Huang2022,Wilary2021}. Notable examples such as ChemDataExtractor\cite{Swain2016,Mavracic2021} and MatSciBERT\cite{Gupta2022} demonstrate the feasibility of extracting materials entities and associated properties from text using rule-based systems and domain-trained language models. However, such approaches typically require substantial manual effort, such as developing parsing rules and curating large human-annotated corpora for model training or fine-tuning\cite{Kononova2021,Polak2023a,Jiang2025}, which limits their adaptability and scalability across different materials systems, properties, and reporting styles.

The advent of large language models (LLMs) unlocks new possibilities for automated data extraction. Unlike traditional NLP approaches, LLMs are pre-trained on massive and diverse corpora, enabling them to capture context, reconcile variations in terminology, and adapt to heterogeneous reporting formats across the literature\cite{Schilling-Wilhelmi2024,Miret2025,Jiang2025}. These capabilities make LLMs particularly well-suited for extracting complex, unstructured materials data. Nevertheless, existing applications in materials science have primarily focused on text-based extraction tasks\cite{Dagdelen2024,Polak2023,Lee2024,Lee2023a,Zhang2024c,Ansari2023,Gupta2024,Yang2025,Rihm2025,Shi2025,Wei2025,Zheng2023,Sipila2025}, with only limited efforts exploring tabular data extraction\cite{Odobesku2025,Kang2025,Itani2024,Polak2023a,Circi2024,Mahjoubi2024}. Their broader potential, especially for integrating information scattered across multiple textual passages and tables and for handling diverse materials systems, remains largely untapped. This gap is particularly critical for composite materials like concrete, where material composition is defined not by a single chemical formula (as in many other materials systems), but by mixtures of multiple constituents, each characterized by distinct chemical and physical attributes that are reported in a heterogeneous and fragmented manner.

\begin{figure}[!htb]
\centering
\includegraphics[scale=0.9]{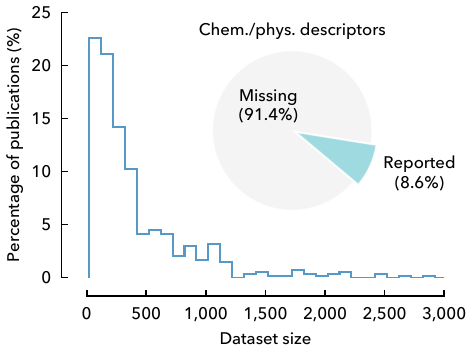}
\caption{\textbf{Survey of dataset sizes used in peer-reviewed publications applying machine learning on cement and concrete materials.} The survey focuses on laboratory datasets for regression tasks. Dataset sizes after data preprocessing were recorded, with the maximum size taken for studies using multiple datasets. The inset shows the proportion of publications whose datasets include chemical or physical descriptors of binders versus those that lack them. Data source: Web of Science; period: 1990--2024. Detailed survey methodology is provided in Supplementary Note~2.}
\label{fig:literature_analysis}
\end{figure}

Concrete is the most-used manufactured material and is responsible for 8--9\% global anthropogenic CO$_2$ emissions\cite{Monteiro2017}. With its central role in global infrastructure and demand projected to grow by up to 23\% by 2050\cite{InternationalEnergyAgency2018}, the need for more sustainable and durable concrete has never been more urgent. Despite this, the design of concrete materials remains largely empirical and trial-and-error-based\cite{DeRousseau2018,Buffenbarger2023}, leading to suboptimal performance, inefficient resource use, and unnecessary CO$_2$ emissions. Emerging data-driven design frameworks have demonstrated the potential to reduce material costs by $\sim$30\% and embodied CO$_2$ emissions by $\sim$60\% compared to industry-standard designs\cite{Pfeiffer2024}. However, progress in this direction is fundamentally constrained by the missing underlying data foundation\cite{Li2022a,Li2022}. Most datasets used in existing machine learning (ML) studies contain fewer than 500 experimental records\cite{BenChaabene2020,Nunez2021a,Li2022a} (Fig.~\ref{fig:literature_analysis}) and are narrowly focused on specific materials systems or properties; moreover, more than 90\% lack critical materials attributes such as binder chemical composition or physical properties (inset of Fig.~\ref{fig:literature_analysis}). The most widely used open dataset, compiled by Yeh\cite{Yeh1998,Yeh2007b} nearly three decades ago, contains only 1,030 records. While several larger industrial datasets have been explored\cite{Young2019,DeRousseau2019,Zhang2021a,Pfeiffer2024}, they are typically proprietary and not publicly accessible. These limitations are shared across many materials domains and underscore the urgent need for scalable approaches to constructing larger, accessible, and information-rich datasets for data-driven materials innovation.

To address the long-standing data scarcity bottleneck in materials informatics, we present an LLM-powered automated data extraction pipeline that systematically retrieves and structures key materials information from both tables and text in scientific publications. Concrete materials are used as a representative and particularly challenging demonstration case due to their compositional complexity and heterogeneous reporting practices. Specifically, we focus on blended cement systems---one of the most promising strategies for concrete decarbonization\cite{InternationalEnergyAgency2018}---in which portland cement, the dominant source of emissions in concrete manufacturing, is partially replaced with supplementary cementitious materials (SCMs) derived from industrial and municipal wastes or naturally abundant resources\cite{Snellings2012,Juenger2019,Snellings2023} (see Supplementary Note~1 for details). Our pipeline employs a sequential chain of LLM-based agents to extract and integrate composition--process--property attributes scattered across multiple sections of a publication. These attributes include binder properties (chemical composition and physical attributes of raw cementitious materials), mixture proportions, curing conditions (age, temperature, and humidity), specimen dimensions (volume and aspect ratio), and mixture properties (here focusing on compressive strength, one of the most critical performance metrics for construction applications\cite{ACICommittee2112002}). Specialized extraction agents are designed to identify relevant text and tables, as well as acronym definitions, which are often separated from the corresponding data. These are complemented by processing agents that perform data cleaning tasks such as acronym expansion, table format reconciliation, unit normalization, naming standardization, irrelevant information filtering, and record-level merging, ultimately constructing complete and internally consistent materials records.

We benchmark our pipeline across a wide range of open and proprietary LLMs. Most models achieve $F_1$ scores above 0.90 across diverse extraction categories, with the best-performing model reaching 0.98, demonstrating robust pipeline performance. The pipeline has been applied to approximately 280 publications on blended cement systems, screened from an initial corpus of more than 27,000 full-text publications, and automatically extracted over 10,000 structured materials records within one hour. This enables the construction of the largest laboratory database for blended cement concrete to date (nearly 9,000 high-quality records with over 100 attributes after postprocessing), encompassing both conventional and emerging SCMs, including fly ash, blast-furnace slag, silica fume, limestone powder, and calcined clay. ML analyses of the curated database reveal that large, diverse, and information-rich datasets are crucial for improving both in-distribution (ID) predictive accuracy and generalization to out-of-distribution (OOD) domains such as unseen materials systems, thereby enabling more reliable and generalizable data-driven materials modeling. Finally, we discuss common challenges encountered in automated data extraction from the scientific literature and propose data reporting recommendations to support the development of large-scale, machine-readable data infrastructures for materials design and discovery. While evaluated here for concrete materials, the modular design of our approach renders it readily adaptable to other materials systems and scientific domains facing data scarcity, broadening its impact across materials informatics and beyond.

\section{Results}
\subsection{Automated data extraction pipeline}

\begin{figure}[!htb]
\centering
\includegraphics[scale=0.9]{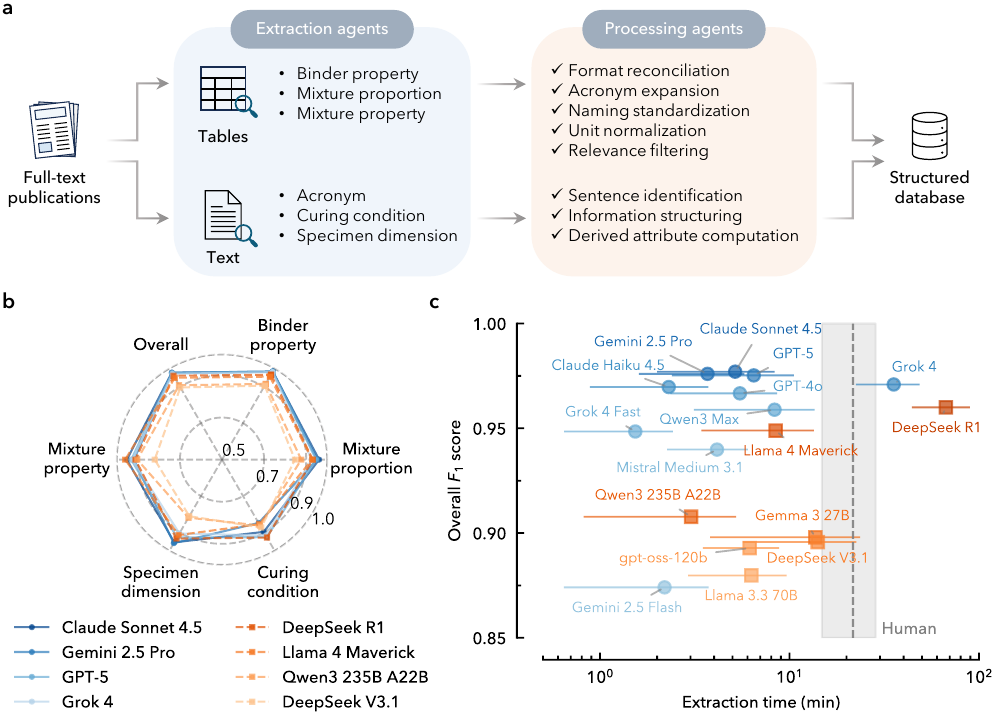}
\caption{\textbf{Overview of the automated data extraction pipeline.} \textbf{a}, The pipeline employs a sequential chain of large language model (LLM)-based extraction and processing agents to retrieve and integrate key concrete materials information from both tables and text in scientific publications, enabling the construction of a structured, machine-readable database. \textbf{b}, \textbf{c}, The pipeline is benchmarked across a range of proprietary (blue) and open (orange) LLMs, evaluated against a human-extracted ground-truth dataset compiled from 58 publications. Panel \textbf{b} shows representative models and their category-wise performance ($F_1$ score), while panel \textbf{c} summarizes the overall $F_1$ score versus extraction time for all investigated models. Color saturation indicates the overall $F_1$ score for each model (darker colors represent higher scores). Error bars in panel \textbf{c} represent the standard deviation of extraction time across the 58 publications for each LLM, and the shaded band reflects the corresponding variation for human experts across 27 publications.}
\label{fig:pipeline}
\end{figure}

Our automated data extraction pipeline (Fig.~\ref{fig:pipeline}a) extracts key concrete materials information from scientific publications, including binder properties, mixture proportions, and mixture properties from tables, as well as acronyms and their definitions, curing conditions, and specimen dimensions from text. The pipeline employs two main categories of specialized agents: (i) extraction agents, which identify and extract relevant information from different sections of a publication, and (ii) processing agents, which clean, reconcile, and integrate the extracted information into complete records for each mixture. Each agent is powered by an LLM. Prompt template and examples are provided in Supplementary Note~4. The pipeline is organized as a sequential, modular agent chain that decomposes the complex extraction workflow into manageable subtasks, while allowing individual components to be flexibly tailored to specific extraction needs.

The pipeline starts by processing a corpus of full-text articles in XML or HTML format (see Section `\nameref{sec:methods_article_retrieval}' for details). These structured formats are preferred over PDF files as they explicitly encode document elements (\textit{e.g.}, sections, tables, captions, footnotes) as tagged content. By contrast, PDFs primarily store visual layouts intended for rendering, which makes it difficult to recover the logical structure of the document in a consistent and machine-readable manner\cite{Hong2021,Zhu2022}.

For table extraction, content enclosed by table-related tags (\textit{e.g.}, \texttt{<table>} in XML) is passed to a table extraction agent. Unlike traditional rule-based methods that depend on fixed patterns or publisher-specific markup, this agent leverages the adaptability of LLMs to parse tables with diverse and complex layouts, including multi-level headers, merged cells, and footnotes. This approach ensures robustness to heterogeneous markup conventions across publishers and improves reliability in handling irregular table structures. The tables are parsed into a structured plain-text format, with columns separated by delimiters and rows by line breaks. While more explicit formats such as JSON could be used, plain text offers a more compact representation that reduces token usage and improves scalability for the large, multi-column tables commonly found in concrete materials literature. The parsed tables are then filtered to retain only those containing relevant concrete materials data based on their captions and content, namely binder properties (chemical composition and physical attributes), mixture proportions, and mixture properties (herein compressive strength). 

For the relevant tables, a series of processing agents perform several normalization steps: (i) consolidating multi-level headers into single-level headers by explicitly expanding full column names; (ii) processing footnotes that often define acronyms or provide clarifications and integrating them into the main table; (iii) expanding remaining acronyms into full names using acronym--definition pairs extracted from the text (described below); (iv) filling blank cells that are intentionally left empty when their values are identical to those in preceding rows, thereby preventing potential misalignment errors where values might otherwise shift into incorrect columns; and (v) harmonizing column names to a predefined set of target names (\textit{e.g.}, `Cement', `Fly ash', `Water', `Compressive strength') and standardizing units (\textit{e.g.}, converting lb/yd$^3$ to kg/m$^3$ and psi to MPa) to ensure consistency across tables from different publications, while removing irrelevant columns.

For binder properties, information is sometimes split across multiple tables within a single publication---for example, one table reporting chemical properties and another reporting physical properties, or separate tables for different raw materials. In such cases, additional merging is performed by an agent to consolidate the relevant information into a unified table for consistent downstream processing. For mixture proportions, water content may be reported explicitly in mass per unit volume (\textit{e.g.}, kg/m$^3$) or implicitly \textit{via} the water-to-binder ratio. In the latter case, a dedicated processing agent calculates the water content from the total binder content in each mixture. Similarly, the dosage of superplasticizer (also known as high-range water-reducing admixture) may appear in diverse formats, \textit{e.g.}, mass per unit volume (\textit{e.g.}, kg/m$^3$), volume per unit volume (L/m$^3$), or as a percentage of binder mass. For the latter two formats, a processing agent converts volume-based values to mass-based equivalents or calculates the dosage from the binder content, respectively, ensuring all data are extracted on a consistent basis.

In addition to tables, key information is often reported in the text, such as acronyms (commonly used in tables but not always defined in captions or footnotes) and their definitions, specimen dimensions (geometry and size), and curing conditions (temperature and humidity; note that curing ages are usually reported alongside strength values in strength tables). To extract this information, the pipeline leverages both rule-based methods and LLM-based agents. For instance, regular expressions are used to identify acronyms enclosed in parentheses (\textit{e.g.}, `fly ash (FA)'), thereby constructing a dictionary of acronym--definition pairs that can facilitate table processing. For curing conditions and specimen dimensions, regular expressions are applied across the full text to detect candidate sentences that likely contain relevant information, \textit{i.e.}, sentences containing terms related to curing (\textit{e.g.}, `curing', `temperature', `humidity', `demolded', `moist', or symbols like `$^\circ$C' and `$\pm$') or to specimen geometry and size (\textit{e.g.}, `cylinder', `cube', `length', `diameter', or units such as `mm' and `cm'), respectively. These candidate sentences, along with their preceding and succeeding sentences for context, are then passed to LLM-powered processing agents, which (i) select the most relevant sets of sentences and (ii) identify and structure the targeted information and, for specimen dimensions, compute derived values such as aspect ratios and specimen volumes.

Finally, for each publication, the extracted and processed information from both tables and text is integrated by downstream processing agents to construct complete records for every mixture, encompassing binder properties, mixture proportions, curing conditions, specimen dimensions, and mixture properties. Where applicable, mixture IDs are used to link information from different tables (\textit{e.g.}, mixture proportions and mixture properties). Additional checks by an agent are performed to ensure that the extracted values are physically plausible (\textit{e.g.}, compressive strength must be positive). The finalized records are then aggregated into individual machine-readable datasets (\textit{e.g.}, CSV files) for each publication, which are subsequently combined into a single comprehensive database.

Overall, the extraction and processing agents operate in a sequential chain, with intermediate outputs progressively refined and passed downstream until curated mixture records are obtained. Many of the extraction and processing steps described above---such as interpreting heterogeneous and irregular table structures, reconciling information dispersed across tables and text, and normalizing units and naming conventions---require nuanced contextual understanding and domain expertise that are particularly challenging to implement using traditional deterministic, rule-based methods, given the variability and complexity of scientific reporting. The use of LLM-powered agents makes these operations feasible in a flexible and context-aware manner, substantially improving the robustness and accuracy of the extraction process. Combined with the modular pipeline design, which allows individual agents to be readily modified or extended, this framework directly addresses long-standing bottlenecks in concrete data curation and provides a scalable foundation for constructing large, high-quality, machine-readable data infrastructure to support data-driven materials discovery.

\subsection{Performance evaluation} \label{sec:results_pipeline_performance}

After analyzing more than 27,000 full-text publications on blended cement systems, we have identified a corpus of 278 publications that are highly relevant and suitable for data extraction (see Section `\nameref{sec:methods_article_retrieval}' for details). From this corpus, 58 publications ($\sim$20\%) were randomly selected, ensuring diversity across journals and publishers, and manually extracted to construct a ground-truth dataset for benchmarking. This dataset consists of 3,015 records, each containing the target information intended for extraction by the pipeline. The pipeline outputs were then compared with these human-extracted ground-truth records to assess extraction performance.

We benchmarked our pipeline using 17 recent open and proprietary LLMs (Fig.~\ref{fig:pipeline}b, c), with each evaluation using the same LLM consistently across all agents. These models were selected to cover flagship and speed-optimized variants across diverse developers, representing a broad range of model scales and capabilities and enabling a comprehensive assessment of extraction performance. The performance of each model was evaluated in terms of $F_1$ score across five categories of target information, namely binder property, mixture proportion, curing condition, specimen dimension, and mixture property. Overall performance was computed by aggregating the results across all categories. 

Fig.~\ref{fig:pipeline}b shows a selection of leading models and their category-wise performance. Most models achieved strong performance across all categories, with $F_1$ scores achieving 0.90 or higher. Across all 17 evaluated models, the pipeline achieved overall $F_1$ scores above 0.85 (Fig.~\ref{fig:pipeline}c), with 12 models exceeding 0.90 and 8 surpassing 0.95. The highest overall score, achieved by Claude Sonnet 4.5\cite{Anthropic2025}, was 0.98. This level of performance is particularly impressive given the inherent complexity and heterogeneity of the data presented in the literature, and the fact that it was achieved without any domain- or task-specific fine-tuning of LLMs. These results highlight both the effectiveness of our pipeline in extracting diverse types of concrete materials information and its strong generalizability across a wide range of LLMs. Additional evaluation results, including full category-wise precision, recall, and $F_1$ score for all 17 LLMs and separate assessments of numerical value and textual information extraction, are provided in Supplementary Note~4.

Table~\ref{tab:piepline_performance} summarizes the performance of the pipeline powered by GPT-4o\cite{OpenAI2024}, which was among the leading models available during the development of the pipeline and was used for downstream data extraction. GPT-4o achieved consistently high precision, recall, and $F_1$ score across all categories, with an overall value above 0.96 for each metric, demonstrating its strong and reliable extraction performance. Nevertheless, the extraction performance for specimen dimensions (extracted from text) was slightly lower than for other categories, indicating opportunities for further improvement. Manual inspection of the errors reveals that rule-based methods were able to identify a broad set of candidate sentences that included the correct ones. However, the downstream LLM-based processing agents occasionally down-selected inappropriate subsets due to insufficient context, for example, when the extracted information referred to tests other than the compressive strength tests targeted in this study. Increasing the context windows of the sentence sets passed to the LLM-based agents may help alleviate such issues. 

\begin{table}[!htb]
\centering
\caption{\textbf{Performance of the automated data extraction pipeline across different categories of concrete materials information.} Results are shown for the pipeline powered by GPT-4o, which was used for downstream data extraction in this work.}
\begin{tabular}{llll}
\toprule
\textbf{Target information} & \textbf{Precision} & \textbf{Recall} & \textbf{$\bm{F_1}$ score} \\
\midrule
Binder property    & 0.98 & 0.96 & 0.97 \\
Mixture proportion & 0.96 & 0.93 & 0.94 \\
Curing condition   & 0.93 & 0.94 & 0.94 \\
Specimen dimension & 0.92 & 0.87 & 0.89 \\
Mixture property   & 0.98 & 0.94 & 0.96 \\
\midrule
\textbf{Overall}   & 0.98 & 0.96 & 0.97 \\
\bottomrule
\end{tabular}
\label{tab:piepline_performance}
\end{table}

In addition to extraction accuracy, we assessed the efficiency of the pipeline by comparing the average processing time per publication between LLMs and human experts (Fig.~\ref{fig:pipeline}c). Human experts required an average of 21.6 minutes per publication. In contrast, LLM extraction times varied by model due to differences in model architecture and API latency, but the majority of LLMs (13 out of 17) required, on average, less than 10 minutes per publication. The median of the models' average extraction times was 6.2 minutes, and the fastest model completed extraction in 1.5 minutes. GPT-4o, the model used for downstream data extraction in this work, processed an article in 5.5 minutes, representing a fourfold speedup relative to human experts. 

Further efficiency gains can be achieved by parallelizing the processing of publications, since each article can be handled independently and the associated LLM requests can be submitted concurrently. For example, extracting data from all 278 publications in our curated corpus required less than one hour using GPT-4o with parallelization, compared with approximately 25 hours if processed sequentially. This level of scalability stands in sharp contrast to manual extraction, which may be manageable on a small scale but becomes prohibitively tedious and time-consuming when extended to hundreds or thousands of publications. As such, this LLM-powered pipeline offers a practical and efficient solution for large-scale literature mining and data infrastructure construction. Moreover, it allows flexible adjustment of the extraction scope (\textit{e.g.}, extracting additional properties or experimental details) without the need to re-examine all publications manually. A brief cost analysis is also provided in Supplementary Note~4.

\subsection{Overview of the curated database}

Applying our pipeline with GPT-4o to the curated corpus of 278 publications, we extracted a total of 10,313 records. To ensure data quality and consistency for subsequent analyses and ML modeling, additional cleaning steps were performed to remove duplicate entries and mixtures outside the scope of this study (\textit{i.e.}, those involving non-targeted SCMs) (see Section `\nameref{sec:methods_data_postprocessing}' for details), resulting in a final database of 8,979 unique, high-quality records. This database is distinct in its scale, diversity, and curation quality.

\subsubsection{Scale}

The limited size of existing datasets has long been a critical bottleneck in data-driven concrete research (Fig.~\ref{fig:literature_analysis}). A key feature of our database is its unprecedented scale. For reference, we compare our database with the 10 largest existing laboratory datasets for blended cement concrete strength reported in the literature (the five largest are shown in Fig.~\ref{fig:database}a, with the full list provided in Supplementary Note~5). To the best of our knowledge, the dataset from Jiang \textit{et al.}\cite{Jiang2024} is the previously largest reported in the literature, containing 5,026 records, although it has not been released publicly. By contrast, our curated database of 8,979 unique records---from over 10,000 initially extracted records using our automated pipeline---is nearly twice as large, making it by far the largest openly available dataset in this domain. Other datasets are generally much smaller, and most are not openly accessible or released alongside their publications. The previously largest openly available dataset contains 2,171 records from Imran \textit{et al.}\cite{Imran2023}, whereas the most widely explored dataset in the field, compiled by Yeh in 1998\cite{Yeh1998} and released in 2007\cite{Yeh2007b}, contains only 1,030 records.

\begin{figure}[!htb]
\centering
\includegraphics[scale=0.9]{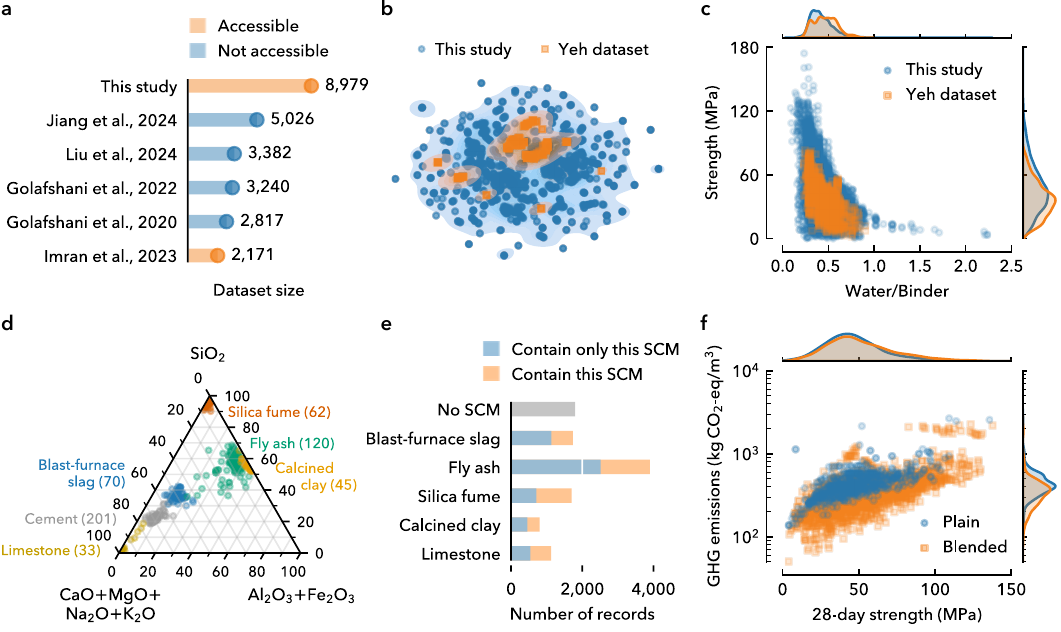}
\caption{\textbf{Overview of the curated database.} \textbf{a}, Comparison of size and accessibility between the database developed in this study and the five largest laboratory compressive strength datasets for blended cement concrete reported in the literature (refs.~\cite{Jiang2024,Liu2024e,MohammadiGolafshani2022,Golafshani2020,Imran2023}). The detailed survey methodology and the comparison with the ten largest datasets are provided in Supplementary Notes~2 and 5, respectively. \textbf{b}, Two-dimensional Uniform Manifold Approximation and Projection (UMAP)\cite{McInnes2020} visualization of the materials feature space of our curated database compared with the Yeh dataset\cite{Yeh1998,Yeh2007b}, the most widely studied open concrete dataset. \textbf{c}, Comparison of data distributions for water-to-binder ratio versus compressive strength between our curated database and the Yeh dataset. Comparisons for other key attributes are provided in Supplementary Note~5. \textbf{d}, Ternary diagram showing the chemical compositions (in wt\%) of the extracted raw cementitious materials. Values in parentheses indicate the number of unique materials. \textbf{e}, Distribution of records across different cementitious systems. \textbf{f}, Greenhouse gas (GHG) emissions of plain portland cement and blended cement mixtures in the curated database. Calculation details are provided in Supplementary Note~5. For panels \textbf{c} and \textbf{f}, marginal kernel density estimates are shown along each axis to represent the data distribution. SCM, supplementary cementitious material.}
\label{fig:database}
\end{figure}

\subsubsection{Diversity}

As discussed earlier, existing datasets are generally small and narrowly focused on specific materials systems, often assembled from a limited number of publications. For example, the blended cement concrete dataset from Yeh\cite{Yeh1998,Yeh2007b} involves only two SCMs (namely, fly ash and blast-furnace slag) and was compiled from only 17 publications, while that from Imran \textit{et al.}\cite{Imran2023} also focuses on these two SCMs, drawing from 15 sources. By contrast, our database encompasses a much broader range of materials systems and was compiled from a significantly larger set of publications (278 in total). It includes five different SCMs, covering both conventional SCMs such as fly ash, blast-furnace slag, silica fume, and limestone powder, as well as emerging SCMs like calcined clay. This enables a wide variety of blended systems, ranging from binary and ternary blends to quaternary blends (\textit{cf.}~Supplementary Note~5). Altogether, the database contains 3,121 unique mixtures and spans a substantially broader materials feature space than existing datasets (Fig.~\ref{fig:database}b).

Additionally, the attribute distributions in our database provide a substantially broader range of values. Notably, the curated database includes mixtures covering a wide range of water-to-binder ratios (0.09--2.2; Fig.~\ref{fig:database}c), aggregate-to-binder ratios, as well as high cement replacement levels (Supplementary Note~5). Such mixtures are rarely captured in smaller datasets but are essential for understanding the full spectrum of mixture performance and enabling reliable data-driven concrete design.

A defining feature of our database is the inclusion of detailed chemical and physical information on the raw cementitious materials used in each mixture, \textit{i.e.}, oxide compositions (\textit{e.g.}, CaO, SiO$_2$, Al$_2$O$_3$, Fe$_2$O$_3$), loss on ignition, specific gravity, and Blaine fineness. Such granularity is critical for assessing the influence of binder characteristics on mixture performance\cite{Li2024}, which remains poorly understood due to the scarcity of such data in existing datasets\cite{Li2022a} (see inset of Fig.~\ref{fig:literature_analysis}). Our database also captures a wide range of unique cementitious materials with diverse chemical compositions (Fig.~\ref{fig:database}d), with their usage broadly distributed across the dataset (Fig.~\ref{fig:database}e).

To further illustrate the diversity of mixture designs captured in the database, we calculated their greenhouse gas emissions (Fig.~\ref{fig:database}f; see Supplementary Note~5 for details). These span a wide range, from 51 to 2,666 kg\,CO$_2$-eq/m$^3$, reflecting the broad design space represented in the extracted data. On average, blended cement systems exhibit lower emissions than plain portland cement mixtures (by approximately 20\%), confirming their potential for emissions reduction. However, substantial variability in emissions persists for both systems at a given strength, with some blended cement mixtures exhibiting emissions comparable to, or even exceeding, those of plain portland cement mixtures. This highlights the tremendous potential of optimizing mixture proportions---\textit{e.g.}, through data-driven design---to fully realize the decarbonization potential of both cementitious systems. Additional analyses of the environmental impact and cost of the mixtures in the curated database are presented in Supplementary Note~5.

\subsubsection{Quality}

The curated database has undergone multiple rounds of cleaning and validation to ensure high data quality and internal consistency. In particular, we systematically inspected extreme values for each feature and computed both the total mass of mixture constituents and the water-to-binder ratio for every record, verifying that the extracted values are physically plausible and consistent with domain knowledge. We also cross-checked the extracted values against the original publications for a randomly selected subset of records to confirm their accuracy. In total, more than 120 publications were verified, together with the 58 publications manually extracted for benchmarking, collectively covering over 7,500 records (more than 80\% of the final database). These validation procedures provide strong confidence in the quality of the curated database. The high overall $F_1$ score (0.97 with GPT-4o) achieved by our extraction pipeline (Table~\ref{tab:piepline_performance}) further supports its reliability.

\subsection{ML analyses}

The composition--process--property relationships of concrete are highly complex and remain difficult to capture comprehensively with existing mechanistic models, especially for blended cement systems where diverse SCM types and attributes dramatically expand the design space. As a demonstration of the utility of our curated database, we performed ML analyses to predict the compressive strength of blended cement concrete from mixture proportions, binder properties, curing conditions, and specimen dimensions. We evaluated a representative set of ML models, including tree-based models (XGBoost (XGB)\cite{Chen2016a}, random forest\cite{Breiman2001}, and LightGBM\cite{Ke2017}), multilayer perceptron\cite{Rumelhart1986}, and support vector machine\cite{Cortes1995}, along with linear regression as a baseline. Model performance was evaluated using root mean square error (RMSE) as the primary metric. Since our curated database uniquely provides detailed chemical and physical descriptors of the raw cementitious materials and contains a significantly larger number of experimental records than existing datasets, we examined two practically important questions: (i) how incorporating these descriptors as additional input features affects model performance, and (ii) how training data size influences both ID accuracy and OOD generalization.

\begin{figure}[!htb]
\centering
\includegraphics[scale=0.9]{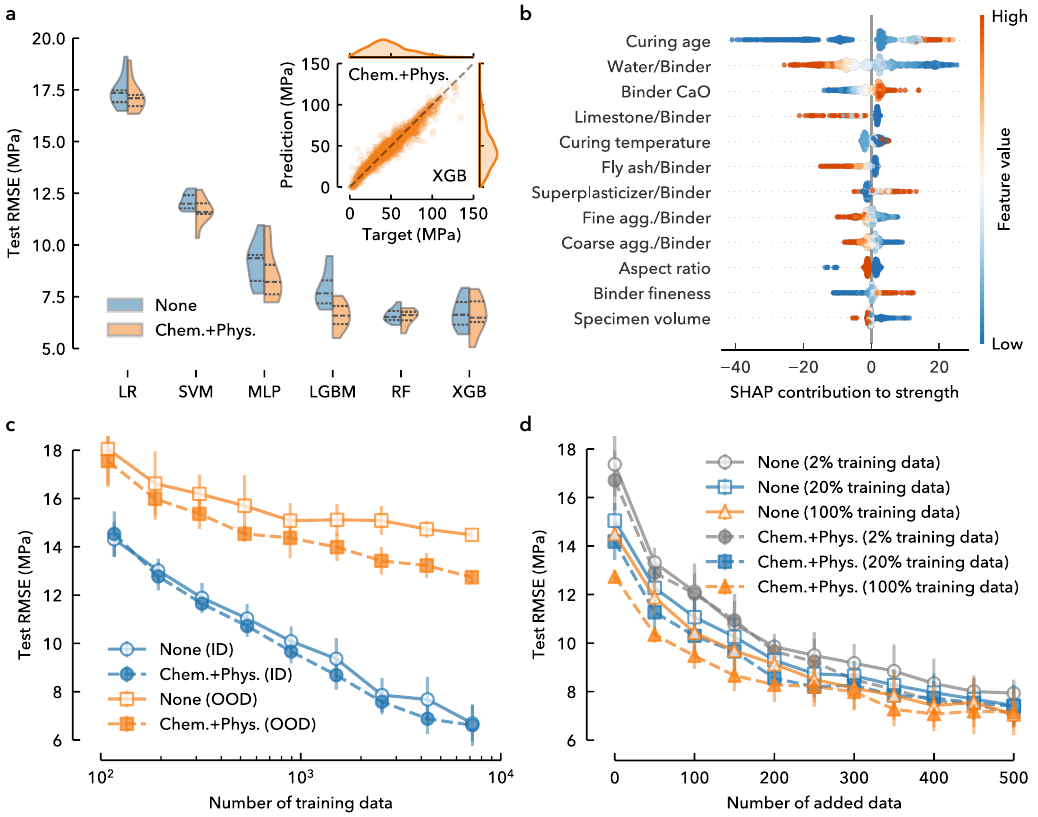}
\caption{\textbf{Machine learning analyses of the curated database.} \textbf{a}, Violin plot showing the in-distribution (ID) performance of models with and without additional chemical and physical descriptors included as input features. The base input features comprise mixture proportions, curing conditions, and specimen dimensions. Each violin includes individual data points from 10 runs with different random seeds, with horizontal dashed lines indicating the median and interquartile range. The inset shows a parity plot of predicted versus measured compressive strength for XGBoost (XGB) using both chemical and physical descriptors as input features, with marginal kernel density estimates shown along each axis to represent the data distribution. \textbf{b}, Summary plot of feature importance, showing the distribution of SHapley Additive exPlanations (SHAP) values for the 12 most important input features of the XGB model with additional chemical and physical descriptors. Feature importance results for the complete set of input features are provided in Supplementary Note~6. \textbf{c}, ID and out-of-distribution (OOD) performance of XGB as a function of training data size, evaluated on a fixed held-out test set. The OOD task involves predicting the compressive strength of ternary and quaternary blended cement mixtures using only data from plain portland cement and binary blended systems for training. \textbf{d}, OOD performance of XGB when varying amounts of test-domain data are added to the training set, with results shown for different fractions of the base training data. For panels \textbf{c} and \textbf{d}, results are averaged over 10 runs with different random seeds, and error bars indicate the standard deviation; results for other models are provided in Supplementary Note~6. LR, linear regression; SVM, support vector machine; MLP, multilayer perceptron; LGBM, LightGBM; RF, random forest; RMSE, root mean square error.}
\label{fig:ml}
\end{figure}

We first evaluated the ID performance of these models using random 80/20 train--test splits of the database and examined the effect of including chemical and physical descriptors of binders as additional input features. The base input features included mixture proportions (expressed as ratios relative to binder content), curing conditions, and specimen dimensions, while the additional descriptors consisted of five major oxide components (CaO, SiO$_2$, Al$_2$O$_3$, Fe$_2$O$_3$, and MgO) and the Blaine fineness of the binder. For blended cements containing multiple constituents, each chemical or physical descriptor of the binder was computed on a mass-weighted basis, with the oxide content or fineness of each raw material weighted by its contribution to the total binder mass. As shown in Fig.~\ref{fig:ml}a, incorporating these descriptors led to slight to modest performance improvements across all models, with the magnitude varying by model. Feature importance analysis for XGB (the top-performing model) using SHapley Additive exPlanations (SHAP)\cite{Lundberg2017} shows that, in addition to curing age and water-to-binder ratio, the CaO content of binders is among the most influential features (Fig.~\ref{fig:ml}b). Binder fineness also exhibits a non-negligible contribution to the predictions. These findings highlight the value of incorporating chemical and physical descriptors into ML models for concrete property prediction. We note that more advanced feature engineering strategies for these descriptors\cite{Xie2018,Gong2023b,Gong2022a,Pedone2015,Oey2020b} are beyond the scope of this work but may further enhance model performance.

To examine the effect of training data size, we trained the XGB model on varying fractions of the training set and evaluated its ID performance on the held-out test set (Fig.~\ref{fig:ml}c). Results show that predictive accuracy improves approximately following a power-law relationship with training data size, consistent with neural scaling laws observed in deep learning\cite{Kaplan2020,Hestness2017,Cheon2023}. This underscores the critical role of large datasets in achieving high predictive performance. The inclusion of chemical and physical descriptors resulted in comparable performance or slight improvements at different training data sizes, likely due to the superior performance of the XGB model achieved with the base features.

We further assessed OOD generalization using a more challenging task, \textit{i.e.}, predicting the compressive strength of ternary and quaternary blended cement mixtures using models trained only on plain portland cement and binary blended systems (Fig.~\ref{fig:ml}c). This setting evaluates a model's ability to generalize from simpler systems to more complex, previously unseen mixture types, which is practically important, as data for more complex blends are often scarce yet these mixtures may offer superior materials use efficiency and sustainability benefits. As expected, OOD performance was substantially worse than ID performance, with RMSE values of XGB with base features exceeding 14 MPa even when the full training set was used. Nevertheless, increasing training data size and including chemical and physical descriptors both led to noticeable improvements in OOD accuracy, underscoring their importance for model generalization to unseen mixture types.

To further enhance OOD generalization, we explored the effect of incorporating varying amounts of test-domain data (ternary and quaternary mixtures) into the training set, mimicking scenarios where targeted experiments are performed and new experimental data become available for model retraining. As shown in Fig.~\ref{fig:ml}d, adding even a small number of test-domain samples led to marked improvements in OOD generalization. Notably, the inclusion of chemical and physical descriptors reduced the number of additional test samples needed to achieve a given accuracy target. For example, achieving an RMSE of 10 MPa required only about 50 additional test samples when using these descriptors, compared to more than 100 additional samples without them. This is particularly useful as it substantially reduces the experimental efforts required. We also varied the fraction of base training data used and observed that the performance gains by adding test-domain samples were generally consistent; however, the benefits of including chemical and physical descriptors were more pronounced at larger base training set sizes. Overall, these results demonstrate the critical role of large, diverse datasets and informative input features in enhancing ID and OOD performance, highlighting the value of our curated database.

\section{Discussion}

Our automated data extraction pipeline, powered by advanced LLMs, has demonstrated strong performance in extracting diverse and complex concrete materials information from scientific literature, including binder properties, mixture proportions, curing conditions, specimen dimensions, and mixture properties, with an overall $F_1$ score up to 0.98. Leveraging this pipeline, we have constructed a comprehensive database of 8,979 unique records for blended cement concrete compressive strength, which is by far the largest openly available dataset in this domain. This database directly addresses long-standing bottlenecks in data scarcity that have hindered data-driven concrete research. ML analyses further demonstrate the critical role of large, comprehensive datasets in advancing concrete materials informatics and enabling more robust predictive modeling.

\subsection{Challenges in automated data extraction}

Despite the remarkable performance of our extraction pipeline, it is important to recognize the inherent challenges of automated data extraction from the literature. Here, we categorize these challenges into two main types, namely LLM-related and information source-related.

A key LLM-related challenge lies in handling arithmetic calculations and unit conversions. For instance, during development we observed occasional errors in computing water content from water-to-binder ratios and total binder content. Similar issues arose in converting superplasticizer dosages into mass-based units when they were reported in volume per unit volume or as a percentage of binder mass. In the latter case, the LLM sometimes directly multiplied the percentage value by the binder mass without dividing by 100 (\textit{e.g.}, interpreting `0.5\% of binder mass' as `$0.5 \times \text{binder mass}$' rather than `$0.005 \times \text{binder mass}$'). Another example involves errors in unit conversion, such as converting specific surface area values from m$^2$/g to m$^2$/kg (\textit{e.g.}, `0.56 m$^2$/g' incorrectly converted to `5.6 m$^2$/kg' rather than `560 m$^2$/kg'). These challenges have been largely mitigated through a combination of strategies: (i) careful prompt design for the processing agents, explicitly instructing LLMs to perform calculations step by step; (ii) chaining multiple specialized agents to decompose complex tasks into simpler subtasks; and (iii) systematic postprocessing of the extracted database, where extreme values can be readily detected and corrected.

Interpreting complex table structures also remains a challenge for LLMs\cite{Zhao2023a,Wu2025}. While our processing agents have been designed to handle diverse and irregular table layouts, occasional errors still occur, particularly with multi-level headers and merged cells, which could misalign values with incorrect columns. These issues have been mitigated through improved prompt design in combination with systematic postprocessing checks. 

Another limitation of LLMs arises from the maximum output token constraint. When tables are very large (\textit{e.g.}, containing hundreds of records and dozens of features) or when multiple tables need to be consolidated into a single large table for downstream processing, the output may be truncated. To mitigate this, our prompt design explicitly instructs the LLM to avoid truncation and fully utilize the available token budget; in addition, for the final step of assembling all records into a machine-readable dataset for each publication (over 100 features in this work), we leverage rule-based methods. Nevertheless, we anticipate that many of these LLM-related challenges will be further alleviated as model capabilities continue to advance.

In terms of information sources, one major challenge pertains to undefined acronyms. Acronyms are widely used in concrete materials literature, particularly in tables, to denote key information such as mixture constituents. To this end, our pipeline specifically examines table captions and footnotes, as well as the main text, to identify acronym--definition pairs and thereby facilitate table processing. However, it is not uncommon for acronyms to appear without being defined anywhere in the publication. While domain knowledge of common acronym usage can sometimes guide the processing agents, this inevitably introduces ambiguity and potential errors. For example, `FA' may be interpreted as either `fly ash' or `fine aggregate', adding an additional layer of complexity and uncertainty to the extraction task. When acronyms are left undefined, even human experts may struggle to interpret them correctly in some cases. It is therefore strongly recommended that acronyms be properly defined in the publication and, when used in a table, accompanied by clear explanations in the caption or footnotes.

\begin{figure}[!htb]
\centering
\includegraphics[scale=0.9]{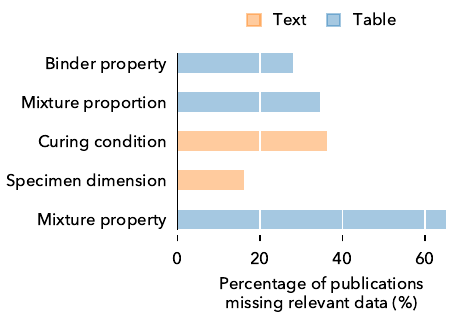}
\caption{\textbf{Overview of missing data in text or tables across different categories of concrete materials information in the literature.} Results for mixture proportion and mixture property are based on the outcome of article screening, while those for binder property, curing condition, and specimen dimension are derived from the curated database (before missing data imputation).}
\label{fig:missing_data}
\end{figure}

Another challenge arises from missing data, which is prevalent in concrete materials literature. According to our survey (see Fig.~\ref{fig:missing_data}), nearly 30\% of the publications in our curated corpus do not report binder properties in tables, while 36\% and 16\% lack explicit reporting of curing conditions and specimen dimensions in the text, respectively---some omitting them entirely, others providing only a reference to the adopted testing standards or earlier studies. Since such information rarely appears in alternative formats, its absence poses a major obstacle for data extraction and subsequent analyses, limiting systematic understanding of its influence on mixture performance. Moreover, references to regional or national standards may not be an ideal substitute for explicit reporting, since standards often encompass multiple specifications and are not always accessible to the broader community. While various imputation techniques can be employed to address missing data for ML modeling, they inevitably introduce additional uncertainty and potential bias. Comprehensive reporting of key materials information is essential to improve data quality and facilitate reliable data-driven research.

The multimodal representation of information presents an additional obstacle. For example, 34\% of the publications do not report mixture proportions in the tables (Fig.~\ref{fig:missing_data}) but may instead describe them in the text, often when mixture design is not the focus of the study and only a limited number of mixtures are considered. Similarly, more than 60\% of the publications do not report mixture properties (herein compressive strength) in tables, which are most likely to be presented in figures. This highlights the importance of multimodal data extraction, which requires careful design of the extraction pipeline to effectively extract and integrate information from different sources and formats. We note, however, that multimodal information extraction, especially from figures, remains a challenging task\cite{Odobesku2025,Zheng2024a,Leong2025,Polak2025,Alampara2025} and is an active area of research in the field of computer science\cite{Liu2023f,Masry2023,Meng2024,Xia2024,Huang2025a}.

Lastly, challenges also arise from data reporting errors in the source publications. Such errors can take multiple forms. One common issue involves incorrect or inconsistent units. For example, a specific surface area value reported in m$^2$/g may, based on its magnitude, actually correspond to m$^2$/kg. Typographical mistakes are another source of error, such as discrepancies between table contents and the corresponding captions or references in the text. In addition, data entry errors can occur, where values reported for one material are mistakenly assigned to another, a problem that can only be detected through domain knowledge and cross-checking with related information. Such issues are problematic not only for automated extraction pipelines but also for human readers, as they introduce uncertainty and risk propagating into downstream tasks if left unchecked. While our curated database has undergone rigorous postprocessing and anomaly detection to flag such inconsistencies, these errors originate in the source literature and cannot always be fully resolved automatically. This highlights the importance of careful data reporting and thorough proofreading by authors, as well as the need for community-wide standards to ensure the accuracy and reliability of published information.

While most challenges are discussed in the context of concrete materials, many are also relevant to other materials systems and scientific domains. For a complementary perspective, we refer readers to refs.~\cite{Hira2024a,Schilling-Wilhelmi2024}, which discuss additional challenges in materials information extraction.

\subsection{Limitations and outlook}

Beyond these extraction challenges, several limitations of the present study point to important opportunities for future development. The following discussion focuses on four main aspects: (i) the data modalities considered, (ii) the granularity of the evaluation, (iii) the degree of automation in pipeline optimization, and (iv) the current scope of extracted information and materials systems. Addressing these issues will be important for extending the framework toward more comprehensive and autonomous materials data extraction. 

Our pipeline, focusing on text and table extraction, has successfully produced the largest laboratory database of blended cement concrete compressive strength, comprising nearly 9,000 unique records with more than 100 attributes. Nevertheless, a wealth of additional data remains embedded in figures (Fig.~\ref{fig:missing_data}), the extraction of which, while beyond the scope of this work, is of paramount importance and remains particularly challenging\cite{Huang2025a}. The diversity of figures---spanning variations in resolution, chart types, visual styles, overlapping elements, and labeling conventions---introduces substantial complexity. Unlike tables, which retain a relatively structured format, figures often encode data implicitly through visual representations that need to be digitized, parsed, and mapped to underlying numerical values. Moreover, the integration of information across axes, legends, and annotations requires sophisticated multimodal reasoning. Recent advances in multimodal and vision-based LLMs for figure understanding across the broader scientific literature\cite{Huang2025a,Meng2024,Xia2024,Liu2024,Lu2024} provide encouraging directions, but further efforts tailored to materials science and concrete research are urgently needed. Another untapped source of information is the supplementary information files published alongside research publications, which often contain valuable materials data and experimental parameters. Future work could explore strategies to effectively incorporate information from supporting documents into the extraction pipeline.

The evaluation in this study is designed to assess the end-to-end extraction workflow from full-text scientific publications to structured database records, by comparing the final LLM-extracted records with human-extracted ground-truth records. Such an evaluation directly reflects the effectiveness of the pipeline for database construction. However, it does not fully isolate or quantify the contributions of individual error sources across intermediate extraction and processing steps. A more granular evaluation of these error sources would provide valuable diagnostic insight for future pipeline development.

The extraction and processing agents in our pipeline are powered by advanced LLMs, and the chained, modular architecture enables flexible adjustment of individual agents for targeted improvements or task-specific customization. In the benchmarking presented in this study, the same LLM was used consistently across all agents to enable a controlled comparison among models. In practical deployment, however, different LLMs could be assigned to different agents based on task complexity, model capability, latency, and cost to enable more efficient performance--cost optimization. Additionally, current pipeline optimization relies on manual prompt engineering and workflow design, which, while effective, require domain expertise and iterative human supervision. Moving forward, a more autonomous and dynamic pipeline can be envisioned\cite{Schilling-Wilhelmi2024,Yao2023a,Odobesku2025,Ansari2023}---one that not only orchestrates tool use but also plans, adapts, and continuously improves workflows in response to task complexity and data modality (\textit{e.g.}, text, tables, or figures). Such a system could autonomously invoke specialized agents, reflect on intermediate outputs, and iteratively refine its decisions, enabling scalable, multimodal extraction with greater robustness and minimal human oversight.

This work focuses on compressive strength, yet other properties such as workability, setting time, and durability are also important for construction applications. Future extensions of the pipeline can capture these additional attributes to provide a more comprehensive understanding of composition--process--property relationships. Moreover, the current pipeline, developed as a proof of concept for blended cement concrete, can be readily adapted to other concrete technologies, including alkali-activated, recycled aggregate, 3D-printed, and ultra-high-performance concretes, enabling large-scale data collection and model development across diverse materials systems.

The present database encompasses five of the most common SCMs---fly ash, blast-furnace slag, silica fume, limestone powder, and calcined clay---which are of immediate industrial relevance. Future work can expand coverage to other SCMs, including natural pozzolans, biomass ashes, recycled glass, concrete recycling fines, waste brick or ceramic powders, and metallurgical (\textit{e.g.}, steel and copper) slags, thereby broadening the compositional diversity of the database and capturing a wider range of sustainable material alternatives. While binder properties have been extracted in this work, aggregate characteristics such as mineral type, shape, size distribution, specific gravity, and absorption are not yet included, as they are less often reported in the literature. Given their potential importance in determining mechanical and durability performance\cite{Hansen2021a,Bentz2017}, extending the pipeline to capture such information would further enhance the database and enable more comprehensive modeling of concrete behavior. In addition, this study focused on mixture proportions reported in mass per unit volume, which are typically used for reporting concrete mixtures; as such, the extracted data are most relevant to concrete-level formulations (see Supplementary Note~5). Mixtures reported only in absolute mass or as percentages/ratios without a reference value were not considered for simplicity, although they can be extracted and converted to a unified format with additional information or assumptions. These cases usually correspond to paste or mortar systems. Incorporating such datasets in the future would enable cross-scale modeling and prediction spanning paste, mortar, and concrete systems.

The curated database, which far surpasses existing manually curated datasets in both scale and scope, serves as a high-quality resource for materials exploration and data-driven research advancements. It represents a significant step toward establishing an open-access data infrastructure for concrete materials informatics, directly addressing the long-standing bottleneck of data scarcity. The automated extraction pipeline can be readily applied to newly published literature, allowing for continuous expansion and regular updates of the database. Beyond its primary focus on concrete, this framework contributes broadly to the fields of materials informatics and automated scientific knowledge extraction, and underscores the importance of comprehensive and rigorous data reporting in the research community.

\section{Methods}
\subsection{Article retrieval and screening} \label{sec:methods_article_retrieval}

We conducted a comprehensive literature search using the Web of Science Core Collection database to assemble a large corpus of publications on cementitious materials, with particular emphasis on those involving fly ash, blast-furnace slag, silica fume, limestone powder, and calcined clay. Only research articles published after 2000 were considered, since they are more likely to be available in XML or HTML formats, which are easier to parse than older publications typically available only in PDF files\cite{Kononova2019,Cruse2022,Zhu2022}. The search yielded 46,820 publications, of which 27,830 full texts in XML/HTML format were successfully retrieved using publisher-specific APIs or web scraping, in full compliance with copyright guidelines and publisher approvals.

To refine this corpus, we developed an LLM-assisted deep screening approach (see Supplementary Note~3 for details). In this procedure, articles were sequentially evaluated and retained only if they met the following criteria: (i) relevance to blended cement systems without incorporating special materials (\textit{e.g.}, non-conventional aggregates, fibers, nanomaterials), non-standard curing conditions (\textit{e.g.}, thermal treatments, chemical exposures), specialized concrete types (\textit{e.g.}, reinforced, prestressed, ultra-high-performance, lightweight/heavyweight, foamed, aerated, polymer-based), or specialized applications (\textit{e.g.}, field-scale studies, soil stabilization, heavy metal immobilization) that could significantly alter mixture performance; (ii) inclusion of essential materials data required for extraction, such as mixture proportions and performance outcomes; and (iii) compatibility with the scope of our data extraction pipeline (\textit{e.g.}, mixture proportions reported in mass per unit volume rather than as absolute mass or percentages without a reference value, and property values reported as absolute rather than relative/normalized measures). This step reduced the corpus to 405 publications. Additional manual checks were then performed to remove false positives and ensure high quality and relevance, resulting in a final set of 278 publications for data extraction.

\subsection{LLMs}

We benchmarked a total of 17 representative LLMs, both open and proprietary. The proprietary models included OpenAI's GPT-5\cite{OpenAI2025a} and GPT-4o\cite{OpenAI2024}, Anthropic's Claude Sonnet 4.5\cite{Anthropic2025} and Haiku 4.5\cite{Anthropic2025a}, Google's Gemini 2.5 Pro and Gemini 2.5 Flash\cite{GeminiTeam2025}, xAI's Grok 4\cite{XAI2025} and Grok 4 Fast\cite{Last2025}, Alibaba's Qwen3 Max\cite{QwenTeam2025a}, and Mistral AI's Mistral Medium 3.1\cite{MistralAI2025}. The open models included OpenAI's gpt-oss-120b\cite{OpenAI2025}, Google's Gemma 3 27B\cite{GemmaTeam2025}, Meta's Llama 4 Maverick\cite{Meta2025} and Llama 3.3 70B\cite{Grattafiori2024}, DeepSeek AI's DeepSeek R1\cite{DeepSeek-AI2025} and DeepSeek V3.1\cite{DeepSeek-AI2024a}, and Alibaba's Qwen3 235B A22B\cite{QwenTeam2025}. Model suffixes denote the number of parameters (in billions) where applicable. These models span both flagship and speed-optimized variants across major developers and differ in model architecture, parameter scale, training strategies, and training data breadth, enabling a comprehensive assessment of extraction performance and efficiency. 

All models were evaluated with the temperature set to 0 to ensure deterministic outputs\cite{Schilling-Wilhelmi2024}. For models offering configurable reasoning settings, the reasoning effort was set to the minimal level permitted; Grok 4 does not expose such controls and was evaluated using its default reasoning behavior. All other inference parameters were kept at their default settings. GPT-5 and GPT-4o were accessed \textit{via} the OpenAI API, and all other models were accessed through the OpenRouter API. Detailed model descriptions and access methods are provided in Supplementary Note~4.

\subsection{Evaluations}

We randomly selected 58 publications from the screened corpus and manually extracted the relevant data to serve as ground truth. The performance of the proposed pipeline was assessed using precision, recall, and the $F_1$ score, as given by\cite{Schilling-Wilhelmi2024}:
\begin{gather}
\text{Precision} = \frac{\mathrm{TP}}{\mathrm{TP} + \mathrm{FP}} \label{eq:precision} \\
\text{Recall} = \frac{\mathrm{TP}}{\mathrm{TP} + \mathrm{FN}} \label{eq:recall} \\
F_1 = 2 \cdot \frac{\text{Precision} \cdot \text{Recall}}{\text{Precision} + \text{Recall}} \label{eq:f1score}
\end{gather}
where true positive (TP) is the number of entities extracted by the pipeline that correctly match those in the ground truth (\textit{i.e.}, as extracted by human experts); false positive (FP) is the number of extracted entities that are not matched with any ground-truth entities or correspond to hallucinated values (\textit{i.e.}, entities that do not exist in the ground truth); and false negative (FN) is the number of ground-truth entities that the pipeline fails to extract or extracts incorrectly. Accordingly, precision measures the proportion of correctly extracted entities among all extracted entities, while recall measures the proportion of correctly extracted entities among all entities in the ground truth. The $F_1$ score gives the harmonic mean of precision and recall. The Levenshtein edit distance was used to map each extracted record with the ground truth for entity-level comparison. An extracted entity is considered a correct match with the ground truth if it falls within a 5\% margin of error for numerical values, or is an exact case-insensitive match for text entities.

\subsection{Postprocessing of the extracted database} \label{sec:methods_data_postprocessing}

After automated data extraction using the pipeline, a total of 10,313 records of blended cementitious materials were obtained. Each record contained information on binder properties (oxide composition, including CaO, SiO$_2$, Al$_2$O$_3$, Fe$_2$O$_3$, MgO, K$_2$O, Na$_2$O, SO$_3$, MnO, TiO$_2$, P$_2$O$_5$, Cl; loss on ignition; specific gravity; and Blaine fineness, each extracted for the individual raw cementitious materials), mixture proportions (cement, blast-furnace slag, fly ash, silica fume, limestone powder, calcined clay, other SCMs, water, superplasticizer, coarse aggregate, and fine aggregate), curing conditions (age, temperature, and humidity), specimen dimensions (volume and aspect ratio), and mixture properties (compressive strength). 

To ensure data quality and consistency prior to further analyses, the database underwent several postprocessing steps: (i) checking extreme values for each feature, verifying the total mass of mixture constituents and water-to-binder ratios, and conducting random manual checks with corrections as needed; (ii) removing records containing SCMs other than the five targeted SCMs (\textit{i.e.}, fly ash, blast-furnace slag, silica fume, limestone powder, and calcined clay; analysis of other SCMs will be considered in future work), mixtures with missing strength values (often omitted due to technical issues during testing), duplicate records (typically reported in multiple articles by the same groups of authors), and mixtures with abnormal total mass (likely due to reporting errors); and (iii) imputing missing values using domain knowledge (\textit{e.g.}, assuming specimen dimensions commonly used for paste, mortar, or concrete) or statistical methods (\textit{e.g.}, median imputation for aggregate contents and curing conditions). 

After postprocessing, 8,979 records were retained for subsequent database analyses and ML modeling. The majority of removed records were due to the presence of non-targeted SCMs ($\sim$1,000 records) and duplicate entries ($\sim$260 records).

\subsection{ML models}

We considered several representative ML models, including tree-based models (XGB\cite{Chen2016a}, random forest\cite{Breiman2001}, and LightGBM\cite{Ke2017}), multilayer perceptron\cite{Rumelhart1986}, and support vector machine\cite{Cortes1995}. A linear regression model was included as a baseline. All models were implemented in Python and executed on CPUs.

For the curated database, we conducted 10 independent runs with different random seeds and 80/20 train--test splits to evaluate ID performance. Records sharing identical input feature values but differing output values (\textit{e.g.}, due to repeated measurements used to capture experimental uncertainty) were grouped, and each group was assigned exclusively to either the training or the test set to prevent data leakage. All input features in the training set were standardized to have zero mean and unit variance, with the same transformation applied to the test set. Hyperparameters were optimized using five-fold cross-validation on the training set for each run. The hyperparameter search space was adapted from ref.~\cite{Hollmann2025} and is detailed in Supplementary Note~6. The best combination of hyperparameters was then used to retrain the model on the full training set, and performance was assessed on the held-out test set. Model performance is reported as the mean and standard deviation of the RMSE across the 10 runs. Feature importance was evaluated using SHAP\cite{Lundberg2017}. To examine the effect of training data size, we randomly sampled subsets of the training data for each run and conducted hyperparameter tuning and model training as described above, while keeping the test set unchanged.

For OOD generalization, we designed a task in which all ternary and quaternary blended cement mixtures (with two and three SCMs, respectively) were excluded from training and used exclusively for testing, while models were trained only on plain portland cement and binary blended cement mixtures (with zero and one SCMs, respectively). This setup evaluates the model's ability to generalize from simpler systems to more complex, previously unseen mixture types, a critical capability for practical concrete materials design. Data standardization, hyperparameter tuning, and model training followed the same procedures described above, and results are reported as the mean and standard deviation of the RMSE across 10 independent runs with different random seeds. To mimic scenarios where limited new experimental data are acquired and become available for model retraining, randomly selected subsets of the test-domain samples were added to the training set, and the same samples were removed from the test set. Additionally, the size of the base training set was varied by randomly sampling different fractions of the original training data.

\backmatter

\bmhead{Acknowledgements}

This work was supported by the Department of Civil and Environmental Engineering at Rice University, the Rice Academy Postdoctoral Fellowship, and the Gulf Research Program's Early-Career Research Fellowship. Access to OpenAI's models was primarily supported by the OpenAI Researcher Access Program. Z.L.~thanks Weichao Ying, Diandian Zhao, and Haokai Zhao for early explorations and discussions in natural language processing. During the preparation of the manuscript, the authors used ChatGPT (OpenAI) to assist with language refinement. The authors retained full editorial control over the manuscript and take full responsibility for its final content. Any opinions, findings, conclusions, or recommendations expressed in this work are those of the authors and do not necessarily reflect the views of the supporting organizations.

\bmhead{Author contributions}

Z.L., K.Y., and K.G.~conceived this work. Z.L.~and K.Y.~developed the automated data extraction pipeline. Z.L.~and Q.H.~screened the articles, manually extracted the data for validation, and cleaned the LLM-extracted database. Z.L.~performed the analyses, prepared the visualizations, and drafted the manuscript. K.G.~supervised the research and acquired funding. All authors discussed the results and contributed to the revision of the manuscript.

\bmhead{Competing interests}

All authors are inventors on a pending U.S.~provisional patent application (No.~64/041,318), filed by Rice University on April 16, 2026, which covers the automated data extraction pipeline described in this work. The authors declare no other competing interests.

\bmhead{Data availability}

The database extracted and curated in this study is publicly available on Zenodo at \url{https://doi.org/10.5281/zenodo.18080350}. The human-extracted ground-truth dataset used to benchmark the automated data extraction pipeline is publicly available on Zenodo at \url{https://doi.org/10.5281/zenodo.22132836}. The unified LLM prompt template and representative prompt examples used in the pipeline are provided in the Supplementary Information.

\end{document}


\title[Article Title]{%
  \centering
  \textbf{Supplementary Information}\\[0.5em]
  Large language model-enabled automated data extraction for concrete materials informatics
}

\author[1,2,3]{\fnm{Zhanzhao} \sur{Li}~\orcidlink{0000-0001-7674-7424}}
\author[4]{\fnm{Kengran} \sur{Yang}~\orcidlink{0000-0001-9766-8063}}
\author[1,2,3]{\fnm{Qiyao} \sur{He}~\orcidlink{0009-0005-7496-9126}}
\author*[1,2,3]{\fnm{Kai} \sur{Gong}~\orcidlink{0000-0002-5884-4607}}
\email{\texttt{kg51@rice.edu}}

\affil[1]{\orgdiv{Department of Civil and Environmental Engineering}, \orgname{Rice University}, \orgaddress{\city{Houston}, \state{TX} \postcode{77005}, \country{USA}}}
\affil[2]{\orgdiv{Rice Advanced Materials Institute}, \orgname{Rice University}, \orgaddress{\city{Houston}, \state{TX} \postcode{77005}, \country{USA}}}
\affil[3]{\orgdiv{Ken Kennedy Institute}, \orgname{Rice University}, \orgaddress{\city{Houston}, \state{TX} \postcode{77005}, \country{USA}}}
\affil[4]{\orgname{Independent researcher}, \orgaddress{\city{Princeton}, \state{NJ}, \country{USA}}}

\maketitle
\tableofcontents
\newpage

\section{Blended cement systems}
\label{si:sec:blended_cement_systems}

Cement acts as a hydraulic binder, reacting with water to form a hardened matrix that binds aggregates (\textit{e.g.}, sand, gravel, or crushed stone) and other constituents into a cohesive material. While it is not the largest component of concrete by weight, cement accounts for the majority of concrete-related CO$_2$ emissions (approximately 90\%\cite{Miller2016,Habert2020,Miller2024}). This high carbon footprint arises primarily from the manufacturing of portland cement---the dominant type used worldwide---which involves the decomposition of limestone (releasing CO$_2$) and the use of high-temperature kilning relying on fossil fuels\cite{Miller2024}.

One way to reduce the carbon footprint of cement is to partially replace the CO$_2$-intensive portland cement with supplementary cementitious materials (SCMs), which are often industrial by-products or naturally occurring materials that react either pozzolanically or hydraulically when combined with portland cement in concrete\cite{Snellings2012,Juenger2019,Snellings2023}. This approach creates blended cement systems, in which multiple cementitious components jointly act as binders. Depending on the type and combination of SCMs and the performance requirements of the concrete, the clinker replacement level can reach 50\% or more by mass\cite{Skibsted2019}. By reducing the demand for portland cement clinker, such systems have been identified by the International Energy Agency as one of the most effective pathways to lower CO$_2$ emissions in the concrete industry\cite{InternationalEnergyAgency2018}. Moreover, they enable the valorization of large quantities of industrial and municipal wastes (\textit{e.g.}, fly ash from coal-fired power plants, blast-furnace slag from pig iron production, silica fume from the ferrosilicon industry, municipal solid waste incineration ash, and waste glass, which would otherwise require landfilling and incur associated environmental and economic costs) as well as naturally abundant resources (\textit{e.g.}, natural pozzolan and clay) into high-performance SCMs, thereby enhancing material circularity and promoting sustainability in the built environment. Beyond their environmental benefits, SCMs can also improve key performance attributes of concrete, such as durability and resistance to chemical attack, without significantly compromising strength performance\cite{Thomas2011a,Lothenbach2011}.

As discussed in the main text, the use of blended cement systems introduces additional complexity in the mix design of concrete, as the interactions among multiple cementitious components and other constituents (\textit{e.g.}, aggregates, chemical admixtures) can lead to nontrivial effects on performance. This complexity is further compounded by the inherent variability in the properties of SCMs (\textit{e.g.}, Fig.~\ref{main-fig:database}d of the main text), which can arise from differences in their source materials, production processes, and treatment conditions\cite{Snellings2012}. Due to the limitations of traditional trial-and-error approaches and existing mechanistic models in capturing these complex interactions and variabilities, optimizing mix designs for blended cement systems remains a significant challenge in practice. Consequently, there is a pressing need for data-driven approaches that can effectively capture and model these complex interactions to optimize mix designs for specific performance requirements while minimizing environmental impact.

\section{Literature survey of machine learning studies in concrete science}
\label{si:sec:lit_survey}

The literature survey of machine learning (ML) studies in concrete science followed the procedure developed by Li \textit{et al.}\cite{Li2022a,Li2022}. An initial search was conducted in the Web of Science Core Collection database (accessed in June 2025) using the criteria summarized in Supplementary Table~\ref{si:tab:lit_survey_criteria}, yielding 1,409 publications. Compared with the previous survey by Li \textit{et al.}\cite{Li2022a,Li2022}, which covered the period from 1990 to 2020 and identified 440 publications, the present search extended the time span to 2024. The substantial increase in publication count underscores the rapidly growing interest in applying ML techniques to concrete science.

\begin{table}[!htb]
\centering
\caption{\textbf{Search criteria used for the literature survey of machine learning (ML) studies in concrete science using the Web of Science database (adapted from refs.~\cite{Li2022a,Li2022}).}}
\label{si:tab:lit_survey_criteria}
\begin{tabularx}{\textwidth}{lX}
\toprule
\textbf{Field} & \textbf{Search criterion} \\ \midrule
Language & English \\
Document type & Article \\
Topic$^{a}$ &
Concrete science-related: `concrete*' OR `mortar*' OR `paste*' OR `cement*' \newline AND \newline
ML-related: `artificial intelligence' OR `machine learning' OR `deep learning' OR `neural network*' OR `support vector machine*' OR `random forest' OR `decision tree*' OR `k-nearest' OR `k nearest' OR `k-nn' OR `knn' OR `k-mean*' OR `k mean*' OR `gaussian process*' \\
Research Area &
`Construction \& Building Technology' AND `Materials Science' NOT `Computer Science' \\
Time span & From 1990 to 2024 \\
\bottomrule
\end{tabularx}
\footnotetext{$^{a}$ The asterisks (*) capture plural and variant forms; \textit{e.g.}, `concrete*' matches terms such as `concrete' and `concretes', and `cement*' matches `cement', `cements', `cementitious', `cementing', and related variants.}
\end{table}

To facilitate further screening and analysis, we developed a large language model (LLM)-based literature analysis pipeline\cite{Li2026a} to automatically: (i) identify truly relevant studies (since the initial search may contain irrelevant ones due to broad search criteria) and (ii) extract key metadata from the full texts of the identified publications, including ML task type (regression, classification, or clustering), data source (experimental or computational; laboratory or field), dataset size (recorded after data preprocessing, with the maximum size used for studies employing multiple datasets), and whether the dataset contained chemical or physical descriptors of binders.

As a result, 1,010 out of the 1,409 initial publications were successfully retrieved in full text. Among these, 833 were determined to be relevant, including 556 that employed experimental laboratory data for ML regression tasks. The analysis of these 556 studies is presented in Fig.~\ref{main-fig:literature_analysis} of the main text.

\section{Article retrieval and screening for automated data extraction}
\label{si:sec:article_retrieval}

In this work, we focused on blended cement systems as a demonstration case for automated data extraction using LLMs. Of particular interest were studies involving five specific SCMs, including fly ash, blast-furnace slag, silica fume, limestone, and calcined clay, which are among the most commonly used or promising SCMs in the concrete industry\cite{Scrivener2018,Scrivener2018a}. We performed a broad search in the Web of Science Core Collection database (accessed in May 2025) using the criteria summarized in Supplementary Table~\ref{si:tab:article_retrieval_criteria}. To avoid excluding relevant studies due to the diverse writing styles and terminology used in the literature, we did not further restrict the query with additional keywords such as `supplementary cement* material*', `mineral admixture*', `mineral additive*', `pozzolan*', `blend*', `replace*', `substitut*', `binary', `ternary', or `quaternary'. The search yielded a total of 46,820 publications, of which 27,830 full texts in XML/HTML format were successfully retrieved using publisher-specific APIs or web scraping. As the present study targeted data extraction from text and tables, publications without tables were excluded, resulting in a pool of 25,521 publications for further screening.

\begin{table}[!htb]
\centering
\caption{\textbf{Search criteria used for article retrieval for automated data extraction using the Web of Science database.}}
\label{si:tab:article_retrieval_criteria}
\begin{tabularx}{\textwidth}{lX}
\toprule
\textbf{Field} & \textbf{Search criterion} \\ \midrule
Language & English \\
Document type & Article \\
Topic$^{a}$ &
Concrete science-related: `concrete*' OR `mortar*' OR `paste*' OR `cement*' \newline AND \newline
Supplementary cementitious materials-related: `blast furnace slag*' OR `GGBFS' OR `GGBS' OR `fly ash*' OR `FA' OR `silica fume*' OR `SF' OR `metakaolin*' OR `MK' OR `calcined clay*' OR `limestone*' \\
Time span & From 2000 to 2024 \\
\bottomrule
\end{tabularx}
\footnotetext{$^{a}$ The asterisks (*) capture plural and variant forms; \textit{e.g.}, `concrete*' matches terms such as `concrete' and `concretes', and `cement*' matches `cement', `cements', `cementitious', `cementing', and related variants.}
\end{table}

\begin{figure}[!htb]
\centering
\includegraphics[scale=1]{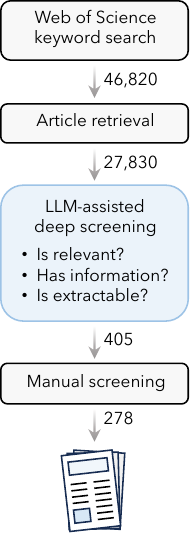}
\caption{\textbf{Overview of the article retrieval and screening steps for automated data extraction.} The numbers along the arrows indicate the number of publications remaining after each step.}
\label{si:fig:article_screening_overview}
\end{figure}

To refine the pool of publications for data extraction, we developed an LLM-assisted deep screening approach\cite{Li2026a} to automatically identify studies that are highly relevant and contain extractable data (Supplementary Fig.~\ref{si:fig:article_screening_overview}). First, we assessed the relevance of each publication to blended cement systems containing at least one of the five targeted SCMs. The title, abstract, and keywords of each publication were evaluated by an LLM-based classifier using structured prompts designed to emulate domain-expert judgment. For each publication, the classifier produced a Boolean decision (`True' or `False' for relevance) accompanied by a brief natural-language explanation summarizing the rationale for its classification. These explanations were used to facilitate random checks by human reviewers to verify correctness and consistency across the screened corpus. Publications involving other SCMs (\textit{e.g.}, natural pozzolans, rice husk ash) were considered relevant and retained if one or more of the above five SCMs were also present, whereas those focusing on any of the following categories were excluded as not relevant, regardless of the presence of the targeted SCMs:

\begin{itemize}
\item Alternative binder systems: Alkali-activated binders, magnesium-based cements, belite cements, calcium aluminate/sulfoaluminate cements, or lime/gypsum-based materials. These systems differ fundamentally in hydration chemistry and are thus outside the scope of portland cement-based systems.
\item Special materials: Unconventional aggregates (\textit{e.g.}, lightweight, heavyweight, recycled, potentially reactive, or marine-origin aggregates, as well as cases where SCMs are used as aggregate replacements), fibers (\textit{e.g.}, steel, glass, polymer), nanomaterials (\textit{e.g.}, nano-silica, graphene, carbon nanotube), nucleation agents (\textit{e.g.}, calcium silicate hydrate seeds), and bio-based or bacterial additives. Such materials can significantly alter the physical properties, hydration behavior, or microstructure of the mixture beyond the scope of conventional blended cement systems.
\item Special curing or exposure conditions: High- or low-temperature curing; steam, autoclave, or microwave curing; and chemical or environmental exposures (\textit{e.g.}, sulfate, chloride, acid, seawater, CO$_2$, freeze--thaw, or fire tests). These specialized curing regimes are usually highly complex and can significantly affect hydration kinetics and microstructural development.
\item Special concrete types: Reinforced, prestressed, precast, ultra-high-performance, lightweight/heavyweight, foamed, aerated, polymer, rubberized, plastic, 3D-printed, repaired, marine, or underwater concretes. These materials introduce confounding factors such as reinforcement, special components, or extreme mix proportions that are not representative of typical blended cement mixtures.
\item Special applications: Field-scale construction studies, soil stabilization, or heavy-metal immobilization, which are often context-specific and may not provide generalizable data for typical concrete mixtures.
\end{itemize}

These exclusion criteria reflect the inherent diversity and complexity of concrete materials research. While the excluded categories represent important and active areas of study, they typically require additional, system-specific descriptors to accurately capture their composition--process--property relationships and therefore pose greater challenges for automated extraction. The present study demonstrates the feasibility of LLM-enabled automated data extraction for typical blended cement systems, whereas future developments of the proposed pipeline, together with continued advancements in multimodal and autonomous LLMs, are expected to extend this framework to more diverse material systems.

This relevance screening step reduced the number of publications to 2,387. Next, we performed a second-level screening to evaluate data quality and determine whether each publication contained essential materials information. Specifically, we examined whether the publications reported (i) mixture proportion data and (ii) performance results (herein compressive strength) necessary for subsequent analysis.

This process was also automated using LLM-based classifiers that analyzed the caption, header, and content of each table within the retrieved full texts. Two separate classification tasks were designed: one for identifying mixture proportion tables and another for identifying compressive strength tables. In each case, the classifier received structured prompts defining relevant keywords and exclusion rules---for example, distinguishing true mixture proportion tables from those reporting design parameters, statistical summaries, or properties unrelated to compressive strength testing, and distinguishing true compressive strength tables from those focused solely on specified strength values, other mechanical properties, or statistical analyses of strength data (\textit{e.g.}, analysis of variance or regression results). Similar to the relevance screening, the classifier produced a Boolean decision for each table, which was then aggregated at the publication level. 

As a result, 1,564 publications were identified as containing mixture proportion tables, and 835 as containing compressive strength tables. Only 610 publications included both types of information and were therefore retained for the subsequent screening stage. 

The third LLM-assisted screening step assessed whether the information contained in each publication was extractable within the scope of our automated data extraction pipeline. We again used LLM classifiers that analyzed each table’s caption, header, and body and returned a Boolean decision for extractability. Specifically, for mixture proportion tables, the classifier evaluated whether mixture constituents were reported in mass per unit volume (\textit{e.g.}, kg/m$^3$, lb/yd$^3$, or clearly indicated variants such as `kg per cubic meter'). Dosages of superplasticizer expressed in percentages or volume units were also accepted. Since the majority of mixture proportion data in the literature follow this format, tables reporting only absolute masses or percentages/ratios without a reference value were not considered extractable in this study, although such data could be extracted and converted to a unified format with additional information or assumptions. For compressive strength tables, the classifier determined whether strengths were reported in absolute values with units (\textit{e.g.}, MPa, N/mm$^2$, psi, ksi). Tables presenting only relative or normalized values were excluded, as their reference strengths are usually reported in text or figures, rendering automated extraction rather challenging.

After this screening step, 453 publications were found to contain extractable mixture proportion tables, 531 contained extractable compressive strength tables, and 405 contained both types of extractable information. Overall, the LLM-assisted deep screening approach reduced the initial pool of 25,521 publications with tables to 405 (Supplementary Fig.~\ref{si:fig:article_screening_overview}). Manual verification was then conducted to remove false positives and ensure that the retained publications were of sufficient quality for automated extraction---for example, excluding those with missing aggregate or binder contents in mixture proportion tables, acronyms undefined throughout the publication, or tables with low readability and ambiguous structure---yielding a final set of 278 publications for data extraction. 

This outcome demonstrates the effectiveness of LLMs in enabling large-scale literature screening for materials informatics while substantially reducing the need for manual effort in the process. While our screening criteria may appear complex and excluded a large number of publications, we emphasize that (i) neither multi-step article screening nor subsequent automated data extraction tasks could be easily achieved using traditional rule-based methods given the diversity of scientific writing and data reporting styles, which was, however, made feasible here by the use of LLMs and (ii) concrete materials research encompasses a wide range of topics, offering many opportunities for future work to extend the present workflow to other material systems and properties toward the development of large data infrastructures for concrete materials design and discovery.

\section{Details of large language models} 
\label{si:sec:llm_details}

\subsection{Models}

All open and proprietary large language models (LLMs) evaluated in this study are listed in Supplementary Table~\ref{si:tab:llm_models}, along with their corresponding API model codes and developers. The open models were accessed through OpenRouter, but they can also be downloaded and deployed locally.

\begin{table}[!htb]
\centering
\caption{\textbf{The specific models used in this study.} GPT-5 and GPT-4o were accessed \textit{via} the OpenAI API, while all other models were accessed through the OpenRouter API.}
\label{si:tab:llm_models}
\begin{tabular}{lll}
\toprule
\textbf{Model name} & \textbf{API model code} & \textbf{Developer} \\
\midrule
\multicolumn{3}{l}{\textit{Proprietary models}} \\
\hspace{2mm}GPT-5\cite{OpenAI2025a}                 
    & \texttt{gpt-5-2025-08-07}                & OpenAI \\
\hspace{2mm}GPT-4o\cite{OpenAI2024}                
    & \texttt{gpt-4o-2024-08-06}               & OpenAI \\
\hspace{2mm}Claude Sonnet 4.5\cite{Anthropic2025}    
    & \texttt{anthropic/claude-sonnet-4.5}     & Anthropic \\
\hspace{2mm}Claude Haiku 4.5\cite{Anthropic2025a}     
    & \texttt{anthropic/claude-haiku-4.5}      & Anthropic \\
\hspace{2mm}Gemini 2.5 Pro\cite{GeminiTeam2025}           
    & \texttt{google/gemini-2.5-pro}           & Google \\
\hspace{2mm}Gemini 2.5 Flash\cite{GeminiTeam2025}          
    & \texttt{google/gemini-2.5-flash}         & Google \\
\hspace{2mm}Grok 4\cite{XAI2025}                   
    & \texttt{x-ai/grok-4}                     & xAI \\
\hspace{2mm}Grok 4 Fast\cite{Last2025}              
    & \texttt{x-ai/grok-4-fast}                & xAI \\
\hspace{2mm}Qwen3 Max\cite{QwenTeam2025a}           
    & \texttt{qwen/qwen3-max}                  & Alibaba \\
\hspace{2mm}Mistral Medium 3.1\cite{MistralAI2025}    
    & \texttt{mistralai/mistral-medium-3.1}    & Mistral AI \\
\midrule
\multicolumn{3}{l}{\textit{Open models}} \\
\hspace{2mm}gpt-oss-120b\cite{OpenAI2025}           
    & \texttt{openai/gpt-oss-120b}             & OpenAI \\
\hspace{2mm}Gemma 3 27B\cite{GemmaTeam2025}         
    & \texttt{google/gemma-3-27b-it}           & Google \\
\hspace{2mm}Llama 4 Maverick\cite{Meta2025}          
    & \texttt{meta-llama/llama-4-maverick}     & Meta \\
\hspace{2mm}Llama 3.3 70B\cite{Grattafiori2024}       
    & \texttt{meta-llama/llama-3.3-70b-instruct} & Meta \\
\hspace{2mm}DeepSeek R1\cite{DeepSeek-AI2025}        
    & \texttt{deepseek/deepseek-r1-0528}       & DeepSeek \\
\hspace{2mm}DeepSeek V3.1\cite{DeepSeek-AI2024a}      
    & \texttt{deepseek/deepseek-chat-v3.1}     & DeepSeek \\
\hspace{2mm}Qwen3 235B A22B\cite{QwenTeam2025}        
    & \texttt{qwen/qwen3-235b-a22b-2507}       & Alibaba \\
\bottomrule
\end{tabular}
\end{table}

\subsection{Prompt template and examples}

For all LLM-powered extraction and processing agents in this study, we employed a unified prompt template (Supplementary Fig.~\ref{si:fig:prompt_template}) consisting of the following components: (i) \texttt{\{System Prompt\}}, which defines the overall role, expertise, and behavior of the LLM-based agent, for example, `You are an expert in cement and concrete science with strong skills in extracting and processing data from scientific literature'; (ii) \texttt{\{Task\}}, which specifies the specific task to be performed; (iii) \texttt{\{Instruction\}}, which provides detailed guidelines and rules for completing the task; (iv) \texttt{\{Examples\}} (optional), which demonstrate representative input--output pairs with explanations to guide the model; and (v) \texttt{\{Output Format\}}, which prescribes the desired structure and format of the output. 

\begin{prompt}[Prompt Template]
\begin{verbatim}
{System Prompt}
{Task}
{Instruction}
{Examples}
{Output Format}
\end{verbatim}
\end{prompt}
\vspace*{-\intextsep}
\begin{figure}[!htb]
\caption{\textbf{Prompt template used in the automated data extraction pipeline.} 
The placeholders in curly brackets (\texttt{\{text\}}) are filled with task-specific content, and the \texttt{Examples} section is optional depending on task complexity.}
\label{si:fig:prompt_template}
\end{figure}

This structured design enables straightforward adaptation of prompts to diverse tasks while maintaining a consistent framework for developing and integrating agents within the automated data extraction pipeline. Representative examples of such prompts are shown in Supplementary Figs.~\ref{si:fig:prompt_example_table_extraction}--\ref{si:fig:prompt_example_text_processing}.

\begin{prompt}[Prompt Example for Table Extraction]
\textbf{System Prompt}\\[-5ex]
\begin{Verbatim}[breaklines, breaksymbol={}]
You are an expert in cement and concrete science with strong skills in extracting and processing data from scientific literature.
\end{Verbatim}
\PromptSep

\textbf{Task}\\[-5ex]
\begin{Verbatim}[breaklines, breaksymbol={}]
Given the following tables, identify and return any table containing information about mixture proportions of concrete, mortar, or paste mixtures:

{parsed_tables} 
\end{Verbatim}
\PromptSep

\textbf{Instruction}\\[-5ex]
\begin{Verbatim}[breaklines, breaksymbol={}]
If multiple tables are relevant, return all of them. Ensure that no information (such as footnotes) is omitted.
\end{Verbatim} 
\PromptSep

\textbf{Output Format}\\[-5ex]
\begin{Verbatim}[breaklines, breaksymbol={}]
Return only the relevant table(s) without any additional text or explanation.
\end{Verbatim}
\end{prompt}
\vspace*{-\intextsep}
\begin{figure}[!htb]
\caption{\textbf{Prompt example for table extraction in the automated data extraction pipeline.} This example illustrates how a large language model is prompted to identify and extract mixture proportion tables from a set of parsed tables. The placeholder \texttt{\{parsed\_tables\}} represents the list of all tables parsed from a publication.}
\label{si:fig:prompt_example_table_extraction}
\end{figure}

\begin{prompt}[Prompt Example for Table Processing]
\textbf{System Prompt}\\[-5ex]
\begin{Verbatim}[breaklines, breaksymbol={}]
You are an expert in cement and concrete science with strong skills in extracting and processing data from scientific literature. You will need to perform arithmetic operations such as addition and multiplication.
\end{Verbatim}
\PromptSep

\textbf{Task}\\[-5ex]
\begin{Verbatim}[breaklines, breaksymbol={}]
Given the following table:

{table}

Perform the calculations to convert the superplasticizer content from percentage to mass per unit volume (kg/m³).
\end{Verbatim}
\PromptSep

\textbf{Instruction}\\[-5ex]
\begin{Verbatim}[breaklines, breaksymbol={}]
1. Identify the 'Superplasticizer (%)' column in the table. It may also appear as 'High-range water-reducing admixture (%)'. The values should be expressed in percentages.
2. Identify the 'Total binder content (kg/m³)' (or 'Total cementitious materials') column. If this column is not present, compute the total binder content by summing the mass per unit volume (kg/m³) of all binder constituents reported in the table.
3. Calculate the amount of superplasticizer (in kg/m³) using the following formula:
   Superplasticizer (kg/m³) = Superplasticizer (%) × Total binder content (kg/m³)
4. Replace the 'Superplasticizer (%)' column with the calculated 'Superplasticizer (kg/m³)' values.
\end{Verbatim} 
\PromptSep

\textbf{Example}\\[-5ex]
\begin{Verbatim}[breaklines, breaksymbol={}]
For example, given this table:

Mix ID | Superplasticizer (%) | Total binder content (kg/m³) | Others
-------|----------------------|------------------------------|-------
1      | 0.32                 | 550                          | -
2      | 0.30                 | 550                          | -
3      | 0.30                 | 500                          | -

The processed table should be:

Mix ID | Superplasticizer (kg/m³) | Total binder content (kg/m³) | Others
-------|--------------------------|------------------------------|-------
1      | 1.76                     | 550                          | -
2      | 1.65                     | 550                          | -
3      | 1.50                     | 500                          | -

Explanation:
- For row 1, the amount of superplasticizer is 0.32% × 550 kg/m³ = 1.76 kg/m³.
- For row 2, the amount of superplasticizer is 0.30% × 550 kg/m³ = 1.65 kg/m³.
- For row 3, the amount of superplasticizer is 0.30% × 500 kg/m³ = 1.50 kg/m³.
\end{Verbatim} 
\PromptSep

\textbf{Output Format}\\[-5ex]
\begin{Verbatim}[breaklines, breaksymbol={}]
Return only the processed table without any additional text or explanation.
\end{Verbatim}
\end{prompt}
\vspace*{-\intextsep}
\begin{figure}[!htb]
\caption{\textbf{Prompt example for table processing in the automated data extraction pipeline.} This example illustrates how a large language model is prompted to convert the dosages of superplasticizer from percentages to mass per unit volume (kg/m$^3$) based on the total binder content. The placeholder \texttt{\{table\}} represents the input table to be processed.}
\label{si:fig:prompt_example_table_processing}
\end{figure}

\begin{prompt}[Prompt Example for Text Extraction]
\textbf{System Prompt}\\[-5ex]
\begin{Verbatim}[breaklines, breaksymbol={}]
You are an expert in cement and concrete science with strong skills in extracting and processing data from scientific literature.
\end{Verbatim}
\PromptSep

\textbf{Task}\\[-5ex]
\begin{Verbatim}[breaklines, breaksymbol={}]
Given the following set of candidate sentences:

{candidate_sentences}

Identify all sentences that describe the curing conditions applied to concrete, mortar, or paste specimens, and return their preceding and succeeding sentences as context.
\end{Verbatim}
\PromptSep

\textbf{Instruction}\\[-5ex]
\begin{Verbatim}[breaklines, breaksymbol={}]
1. The matched sentence must include at least one of the following:
   - Curing temperature (e.g., 'cured at 20°C', 'stored at 23±2°C', 'cured at room temperature')
   - Curing humidity (e.g., 'moist curing', '95% RH', 'fog chamber', 'lime-saturated water', 'water tank', 'covered with plastic sheet', 'sealed after demolding')
2. Include all sentences that mention curing conditions, even if they do not explicitly refer to a test.
3. Exclude sentences that clearly describe non-compressive strength tests (e.g., carbonation or sulfate exposure).
4. When uncertain, include all potentially relevant sentences.
5. Provide the sentences exactly as written, without any modification or additional explanation.
6. If no relevant sentences are found, return: 'No relevant sentences found.'
\end{Verbatim} 
\PromptSep

\textbf{Output Format}\\[-5ex]
\begin{Verbatim}[breaklines, breaksymbol={}]
- Matched Sentence: <sentence describing the curing condition>
- Preceding Sentences: <context sentences immediately before>
- Succeeding Sentences: <context sentences immediately after>
\end{Verbatim}
\end{prompt}
\vspace*{-\intextsep}
\begin{figure}[!htb]
\caption{\textbf{Prompt example for text extraction in the automated data extraction pipeline.} This example illustrates how a large language model is prompted to identify sentences describing curing conditions and their surrounding context from a set of candidate sentences. The placeholder \texttt{\{candidate\_sentences\}} represents the set of sentences preliminarily retrieved using regular-expression keyword matching with curing condition-related keywords.}
\label{si:fig:prompt_example_text_extraction}
\end{figure}

\begin{prompt}[Prompt Example for Text Processing]
\textbf{System Prompt}\\[-5ex]
\begin{Verbatim}[breaklines, breaksymbol={}]
You are an expert in cement and concrete science with strong skills in extracting and processing data from scientific literature.
\end{Verbatim}
\PromptSep

\textbf{Task}\\[-5ex]
\begin{Verbatim}[breaklines, breaksymbol={}]
Given the following set of relevant sentences:

{relevant_sentences}

Extract the curing temperature (°C) and curing humidity (%) mentioned in the text.
\end{Verbatim}
\PromptSep

\textbf{Instruction}\\[-2ex]
\begin{Verbatim}[breaklines, breaksymbol={}]
1. Identify the curing temperature:
   - If a specific temperature is mentioned (e.g., 'stored at 23±2°C', 'kept at room temperature (27±3°C)'), extract the value.
   - If 'room temperature' is mentioned without a numerical value, assume 23°C.
   - If no temperature is given, return 'null'.
2. Identify the curing humidity:
   - If a specific relative humidity value is mentioned (e.g., '95% RH', '100% RH'), extract the value.
   - If specimens are 'cured in water' or 'lime-saturated water', set humidity to 100%.
   - If specimens are subjected to 'moist', 'fog', or 'wet' curing, assume 95% humidity unless otherwise specified.
   - If specimens are 'sealed' or 'covered with a plastic sheet', assume 90%.
   - If no humidity condition is given, return 'null'.
3. Consider both matched and context sentences when determining curing conditions.
4. When extracting values, exclude uncertainty terms such as '±2°C'.
5. Return only the processed curing conditions without additional explanations.
\end{Verbatim} 
\PromptSep

\textbf{Output Format}\\[-5ex]
\begin{Verbatim}[breaklines, breaksymbol={}]
- Curing Temperature (°C): <value>
- Curing Humidity (%): <value>
\end{Verbatim}
\end{prompt}
\vspace*{-\intextsep}
\begin{figure}[!htb]
\caption{\textbf{Prompt example for text processing in the automated data extraction pipeline.} This example illustrates how a large language model is prompted to extract curing temperature and humidity values from a set of relevant sentences. The placeholder \texttt{\{relevant\_sentences\}} represents the set of sentences identified as relevant in the previous text extraction step (\textit{cf.} Supplementary Fig.~\ref{si:fig:prompt_example_text_extraction}).}
\label{si:fig:prompt_example_text_processing}
\end{figure}

\subsection{Additional evaluation results}

\begin{figure}[!htb]
\centering
\includegraphics[scale=1]{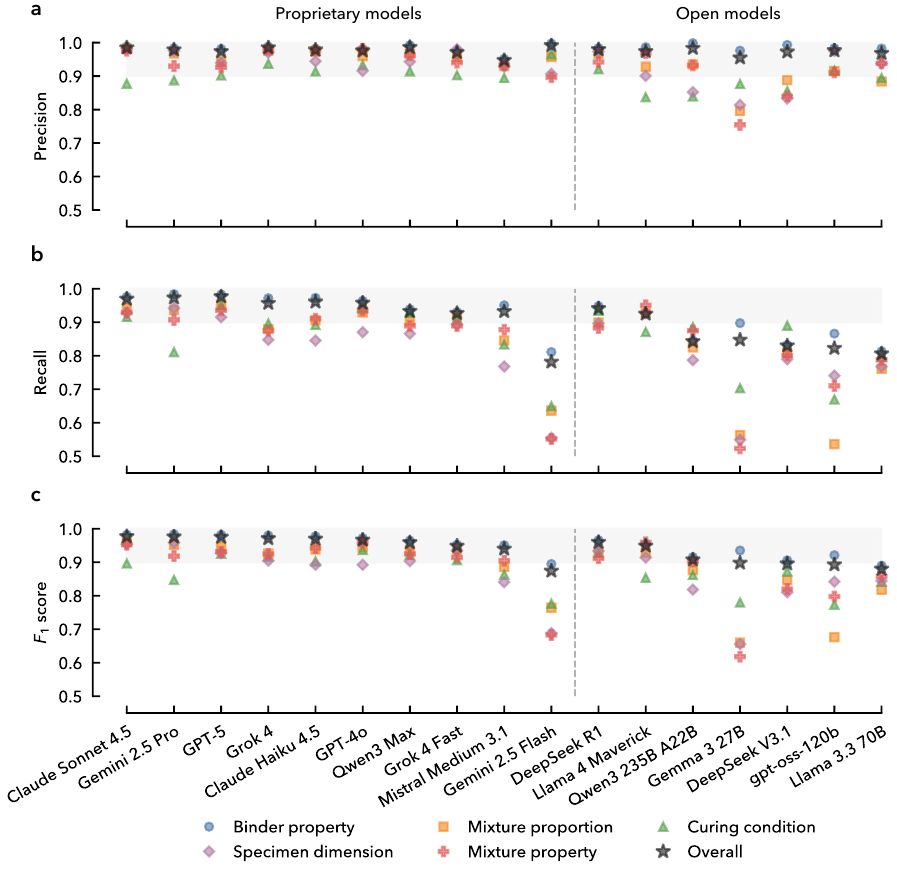}
\caption{\textbf{Performance of the automated data extraction pipeline across different large language models.} \textbf{a}, Precision. \textbf{b}, Recall. \textbf{c}, $F_1$ score. Each marker represents the score for one extraction category or the overall score. The horizontal shaded band indicates scores above 0.90. Proprietary and open models are separated by the vertical dashed line, and models are ranked in descending order of their overall $F_1$ score within each group.}
\label{si:fig:llm_performance}
\end{figure}

Supplementary Fig.~\ref{si:fig:llm_performance} summarizes the detailed performance of the automated data extraction pipeline across all evaluated LLMs in terms of precision, recall, and $F_1$ score. The strong performance achieved by most evaluated models for diverse extraction categories demonstrates the robustness of the proposed pipeline. 

\begin{figure}[!htb]
\centering
\includegraphics[scale=1]{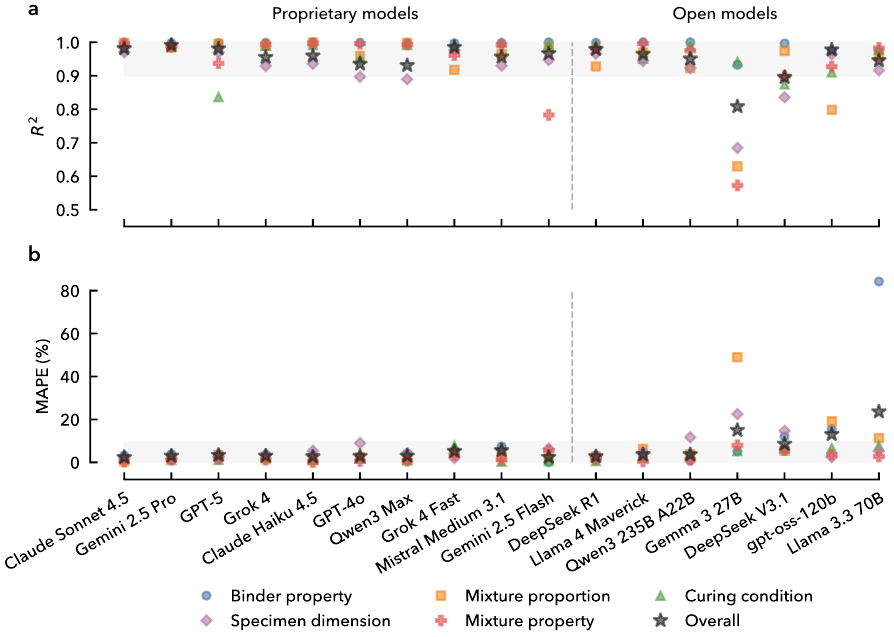}
\caption{\textbf{Performance of the automated data extraction pipeline across different large language models for numerical value extraction.} \textbf{a}, Coefficient of determination ($R^2$). \textbf{b}, Mean absolute percentage error (MAPE). Each marker represents the score for one extraction category or the overall score. The horizontal shaded band indicates $R^2$ above 0.90 or MAPE below 10\%. Proprietary and open models are separated by the vertical dashed line, and models are shown in the same order as in Supplementary Fig.~\ref{si:fig:llm_performance}.}
\label{si:fig:llm_performance_numerical_values}
\end{figure}

\begin{figure}[!htb]
\centering
\includegraphics[scale=1]{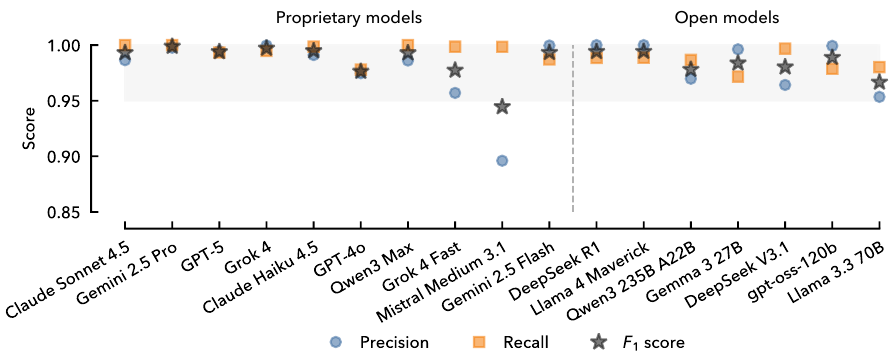}
\caption{\textbf{Performance of the automated data extraction pipeline across different large language models (LLMs) for textual information extraction.} Each marker represents one evaluation metric for a given LLM. The horizontal shaded band indicates scores above 0.95. Proprietary and open models are separated by the vertical dashed line, and models are shown in the same order as in Supplementary Figs.~\ref{si:fig:llm_performance} and \ref{si:fig:llm_performance_numerical_values}.}
\label{si:fig:llm_performance_text}
\end{figure}

To further support the evaluation of the pipeline, we separately assessed its performance for numerical value extraction and textual information extraction (\textit{cf.}~Supplementary Figs.~\ref{si:fig:llm_performance_numerical_values} and \ref{si:fig:llm_performance_text}, respectively). Specifically, regression-based metrics, including the coefficient of determination ($R^2$) and the mean absolute percentage error (MAPE), were used to quantify the magnitude of error between matched extracted and ground-truth numerical values. Among the 17 evaluated models, 15 achieved $R^2>0.90$ and 14 achieved MAPE below 10\%, indicating strong numerical agreement for most models (Supplementary Fig.~\ref{si:fig:llm_performance_numerical_values}). Notably, 9 models further satisfied the stricter criteria of both $R^2>0.95$ and MAPE below 5\%. Although some models, such as Gemma 3 27B and Llama 3.3 70B, exhibited lower performance in certain extraction categories, detailed inspection of the incorrectly extracted data showed that these errors were dominated by extreme or unusual values, which can be readily detected and corrected through postprocessing during database curation. 

The pipeline also showed consistently strong performance for textual information extraction (Supplementary Fig.~\ref{si:fig:llm_performance_text}), with 16 out of 17 models achieving precision, recall, and $F_1$ score all above 0.95. Overall, these results demonstrate the robustness of the proposed pipeline across different LLMs for both numerical value and textual information extraction tasks.

\subsection{Cost analysis}

To evaluate the monetary costs associated with the use of LLMs in the automated data extraction pipeline, we recorded the total number of input and output tokens used for each publication during the extraction of all 278 publications. The token count represents the total number of textual units processed or produced by the model, where a token roughly corresponds to a word, number, punctuation mark, or symbol in the text. On average, processing a publication with GPT-4o required 257,196 input tokens and 52,362 output tokens. According to OpenAI's pricing for GPT-4o as of December 2025, the cost is USD 2.50 per million input tokens and USD 10.00 per million output tokens. Based on these rates, the average processing cost per publication was estimated to be USD 1.17, and the total cost for extracting data from the full set of 278 publications was about USD 324.

For comparison, we estimated the cost of manual data extraction by human reviewers. Assuming an hourly rate of USD 20.00 for domain-expert annotators and an average manual extraction time of 21.6 minutes per publication (as reported in Section `\nameref{main-sec:results_pipeline_performance}' of the main text), the estimated cost per publication is USD 7.20. Thus, manually extracting data from all 278 publications would cost around USD 2,002, which is over six times higher than the cost of using the LLM-powered automated pipeline.

\section{Additional analyses of the curated database}
\label{si:sec:database_additional_analyses}

\subsection{Comparison with existing datasets}

\begin{figure}[!htb]
\centering
\includegraphics[scale=1]{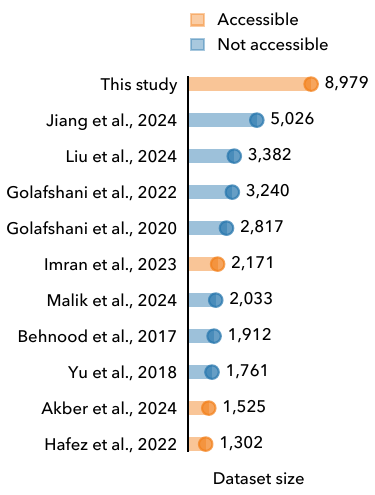}
\caption{\textbf{Comparison of size and accessibility between the database developed in this study and the 10 largest laboratory compressive strength datasets for blended cement concrete reported in the literature (refs.~\cite{Jiang2024,Liu2024e,MohammadiGolafshani2022,Golafshani2020,Imran2023,Malik2024,Behnood2017,Yu2018,Akber2024,Hafez2022}).} The comparison with the five largest datasets is shown in Fig.~\ref{main-fig:database}a of the main text.}
\label{si:fig:dataset_size_comparison_full}
\end{figure}

Supplementary Fig.~\ref{si:fig:dataset_size_comparison_full} provides the full comparison of our curated database with the ten largest laboratory datasets for blended cement concrete reported in the literature, complementing the summary shown in Fig.~\ref{main-fig:database}a of the main text. The substantially larger scale and full accessibility of the present database enable more comprehensive and robust ML analyses than previously possible.

\subsection{Statistical summary}

\begin{figure}[!htb]
\centering
\includegraphics[scale=1]{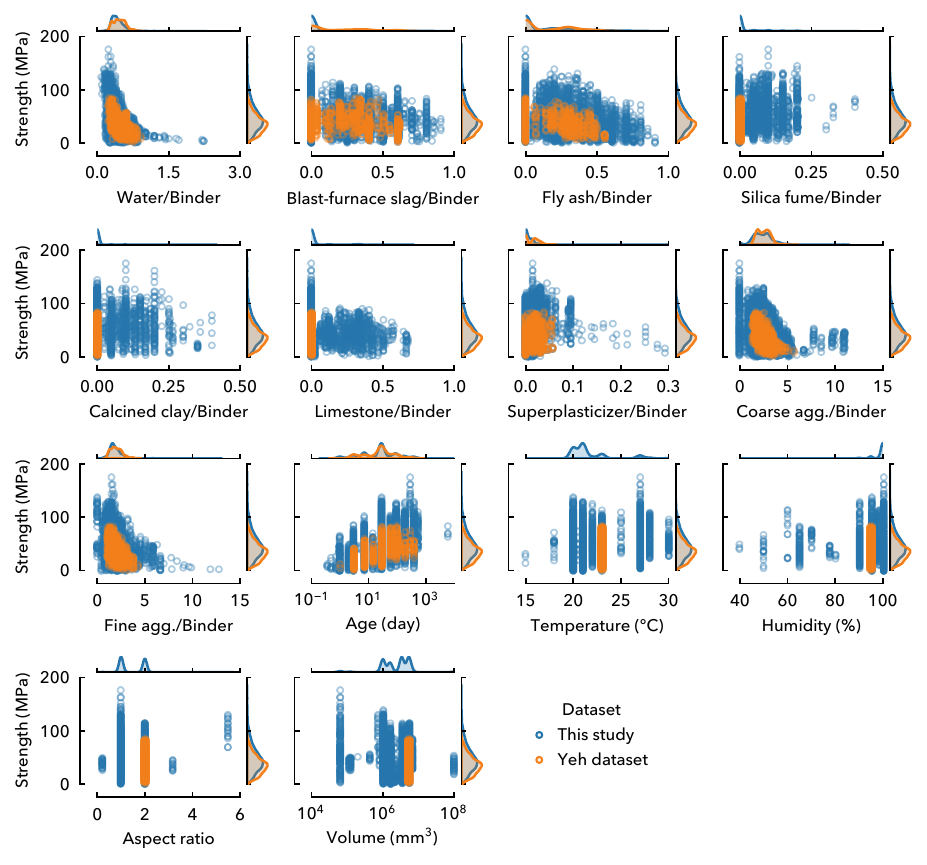}
\caption{\textbf{Relationships between compressive strength and key mixture proportion, curing condition, and specimen dimension features (colored by dataset).} Open circles represent individual data points from the curated database (blue) and the Yeh dataset\cite{Yeh1998,Yeh2007b} (orange). Each subplot shows the bivariate distribution of compressive strength against one feature, with marginal kernel density estimates along each axis indicating the data density. Since curing temperature and humidity were not reported in the Yeh dataset\cite{Yeh1998,Yeh2007b}, constant values of 23$^\circ$C and 95\% relative humidity are assumed here for visualization. agg., aggregate.}
\label{si:fig:dataset_feature_scatter_dataset}
\end{figure}

\begin{figure}[!htb]
\centering
\includegraphics[scale=1]{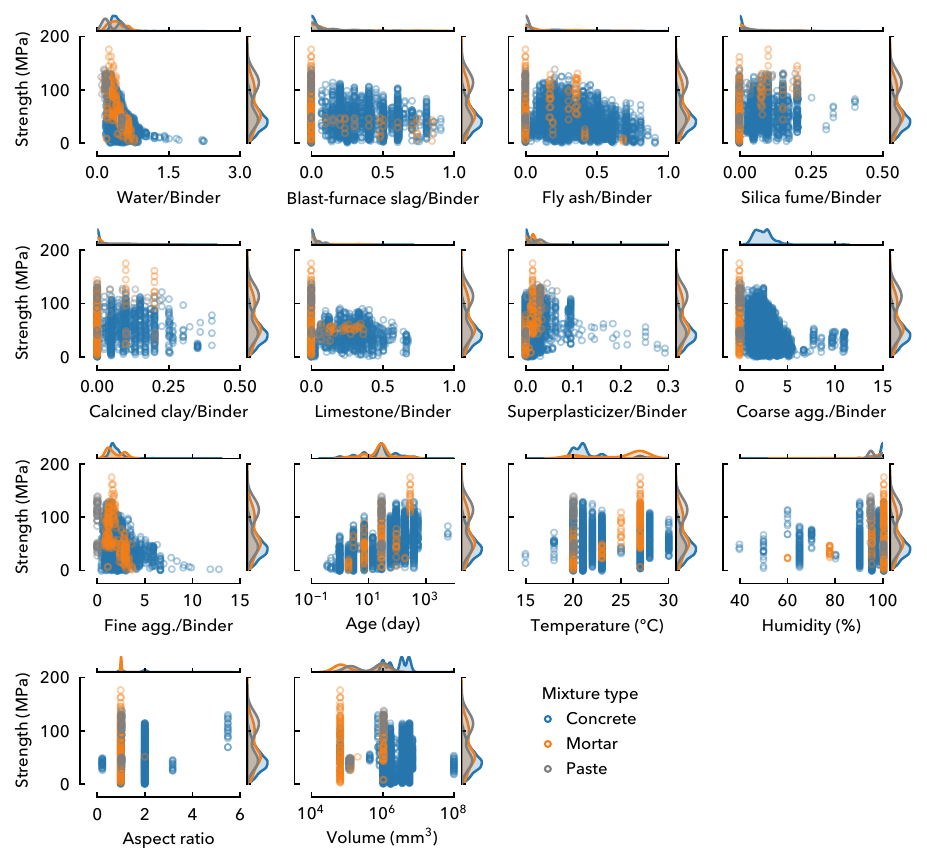}
\caption{\textbf{Relationships between compressive strength and key mixture proportion, curing condition, and specimen dimension features (colored by mixture type).} Open circles represent individual data points for concrete (blue), mortar (orange), and paste (grey) mixtures. Each subplot shows the bivariate distribution of compressive strength against one feature, with marginal kernel density estimates along each axis indicating the data density. agg., aggregate.}
\label{si:fig:dataset_feature_scatter_mixturetype}
\end{figure}

Supplementary Table~\ref{si:tab:dataset_statistical_summary} summarizes key statistical characteristics of the curated database, including mean, standard deviation, minimum, maximum, and interquartile range (25th, 50th, and 75th percentiles) for each attribute. The dataset spans a wide range of mixture proportions, curing conditions, specimen dimensions, and compressive strengths, capturing the variability of blended cement systems reported in the literature. Supplementary Figs.~\ref{si:fig:dataset_feature_scatter_dataset} and \ref{si:fig:dataset_feature_scatter_mixturetype} further show the relationships between compressive strength and some selected key features. Compared with the widely used dataset by Yeh\cite{Yeh1998,Yeh2007b}, the curated database exhibits a substantially broader range of these features, reflecting its larger scale and greater diversity (Supplementary Fig.~\ref{si:fig:dataset_feature_scatter_dataset}). 

\begin{table}[!htb]
\centering
\caption{\textbf{Statistical characteristics of the curated database.}}
\label{si:tab:dataset_statistical_summary}
\begin{tabular}{llllllll}
\toprule
\textbf{Variable}        & \textbf{Mean}      & \textbf{STD}       & \textbf{Min}    & \textbf{25\%}      & \textbf{50\%}      & \textbf{75\%}      & \textbf{Max}        \\
\midrule 
\multicolumn{8}{l}{\textit{Mixture proportion (kg/m$^3$)}} \\
\hspace{2mm}Cement                     & 323.1     & 156.8     & 35.0   & 225.0     & 317.0     & 400.0     & 2,344.0    \\
\hspace{2mm}Blast-furnace slag         & 30.2      & 72.6      & 0.0    & 0.0       & 0.0       & 0.0       & 450.0      \\
\hspace{2mm}Fly ash                    & 60.8      & 89.6      & 0.0    & 0.0       & 0.0       & 103.0     & 544.0      \\
\hspace{2mm}Silica fume                & 9.0       & 24.2      & 0.0    & 0.0       & 0.0       & 0.0       & 469.0      \\
\hspace{2mm}Calcined clay              & 5.3       & 22.1      & 0.0    & 0.0       & 0.0       & 0.0       & 404.7      \\
\hspace{2mm}Limestone                  & 17.0      & 52.4      & 0.0    & 0.0       & 0.0       & 0.0       & 449.0      \\
\hspace{2mm}Water                      & 179.6     & 43.3      & 75.0   & 159.0     & 177.0     & 192.4     & 611.7      \\
\hspace{2mm}Superplasticizer                      & 4.4       & 8.3       & 0.0    & 0.0       & 1.9       & 6.0       & 152.3      \\
\hspace{2mm}Coarse aggregate           & 960.6     & 263.9     & 0.0    & 837.0     & 1,023.7   & 1,124.5   & 1,723.4    \\
\hspace{2mm}Fine aggregate             & 774.3     & 175.4     & 0.0    & 676.0     & 748.0     & 855.0     & 1,816.0    \\
\midrule
\multicolumn{8}{l}{\textit{Mixture proportion (mass ratio)}} \\
\hspace{2mm}Water/Binder               & 0.43      & 0.13      & 0.09   & 0.33      & 0.40      & 0.50      & 2.23       \\
\hspace{2mm}Blast-furnace slag/Binder  & 0.07      & 0.17      & 0.00   & 0.00      & 0.00      & 0.00      & 0.90       \\
\hspace{2mm}Fly ash/Binder             & 0.14      & 0.19      & 0.00   & 0.00      & 0.00      & 0.25      & 0.90       \\
\hspace{2mm}Silica fume/Binder         & 0.02      & 0.05      & 0.00   & 0.00      & 0.00      & 0.00      & 0.40       \\
\hspace{2mm}Calcined clay/Binder       & 0.01      & 0.04      & 0.00   & 0.00      & 0.00      & 0.00      & 0.40       \\
\hspace{2mm}Limestone/Binder           & 0.03      & 0.10      & 0.00   & 0.00      & 0.00      & 0.00      & 0.67       \\
\hspace{2mm}Superplasticizer/Binder               & 0.01      & 0.02      & 0.00   & 0.00      & 0.00      & 0.01      & 0.29       \\
\hspace{2mm}Coarse aggregate/Binder         & 2.44      & 1.20      & 0.00   & 1.69      & 2.33      & 2.98      & 10.94      \\
\hspace{2mm}Fine aggregate/Binder           & 1.90      & 0.76      & 0.00   & 1.50      & 1.76      & 2.18      & 12.74      \\
\midrule
\multicolumn{8}{l}{\textit{Specimen dimension}} \\
\hspace{2mm}Aspect ratio               & 1.47      & 0.53      & 0.24   & 1.00      & 1.00      & 2.00      & 5.45       \\
\hspace{2mm}Specimen volume (mm$^3$)        & 3,296,265 & 5,551,001 & 64,000 & 1,000,000 & 3,375,000 & 5,301,438 & 91,125,000 \\
\midrule
\multicolumn{8}{l}{\textit{Curing condition}} \\
\hspace{2mm}Temperature ($^{\circ}$C)    & 21.7      & 2.3       & 15     & 20        & 21        & 22        & 30         \\
\hspace{2mm}Humidity (\%)       & 98.6      & 4.8       & 40     & 100       & 100       & 100       & 100        \\
\hspace{2mm}Age (day)                  & 59.1      & 143.2     & 0.4    & 7         & 28        & 56        & 5,475       \\
\midrule
\multicolumn{8}{l}{\textit{Mixture property}} \\
\hspace{2mm}Compressive strength (MPa) & 47.2      & 23.8      & 0.9    & 30.4      & 44.3      & 61.0      & 174.4        \\
\bottomrule
\end{tabular}
\vspace*{-\belowrulesep}
\footnotetext{\textit{Note}: STD, standard deviation.}
\end{table}

\subsection{Materials feature space coverage}

In addition to the Uniform Manifold Approximation and Projection (UMAP)\cite{McInnes2020} visualization presented in Fig.~\ref{main-fig:database}b of the main text, we further examined the coverage of the materials feature space using principal component analysis (PCA) and $t$-distributed stochastic neighbor embedding ($t$-SNE)\cite{Maaten2008}, as shown in Supplementary Fig.~\ref{si:fig:dataset_material_feature_space}. These visualizations reinforce the finding that the curated database spans a substantially broader materials feature space compared to existing datasets, such as the widely used dataset by Yeh\cite{Yeh1998,Yeh2007b}.

\begin{figure}[!htb]
\centering
\includegraphics[scale=1]{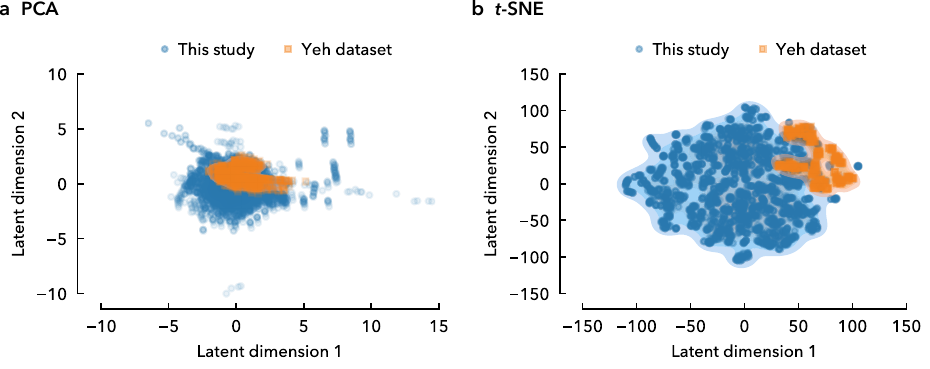}
\caption{\textbf{Two-dimensional visualization of the materials feature space of the curated database and the Yeh dataset\cite{Yeh1998,Yeh2007b}.} \textbf{a}, principal component analysis (PCA). \textbf{b}, $t$-distributed stochastic neighbor embedding ($t$-SNE).}
\label{si:fig:dataset_material_feature_space}
\end{figure}

\subsection{Materials system distribution}

\begin{figure}[!htb]
\centering
\includegraphics[scale=1]{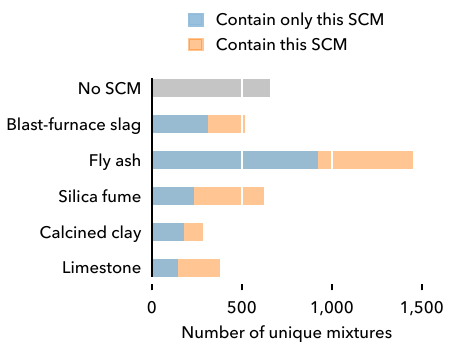}
\caption{\textbf{Distribution of unique mixtures across different cementitious systems in the curated database.}}
\label{si:fig:dataset_scm_distribution_mixture}
\end{figure}

In Fig.~\ref{main-fig:database}e of the main text, we presented the distribution of records across different cementitious systems in the curated database. Here, we further examine the distribution of unique mixtures across these systems in the curated database (Supplementary Fig.~\ref{si:fig:dataset_scm_distribution_mixture}). The results show that the distribution of unique mixtures closely mirrors that of the records, indicating that the data coverage across cementitious systems is not disproportionately influenced by multiple experimental measurements conducted on a limited set of mixtures.

\begin{figure}[!htb]
\centering
\includegraphics[scale=1]{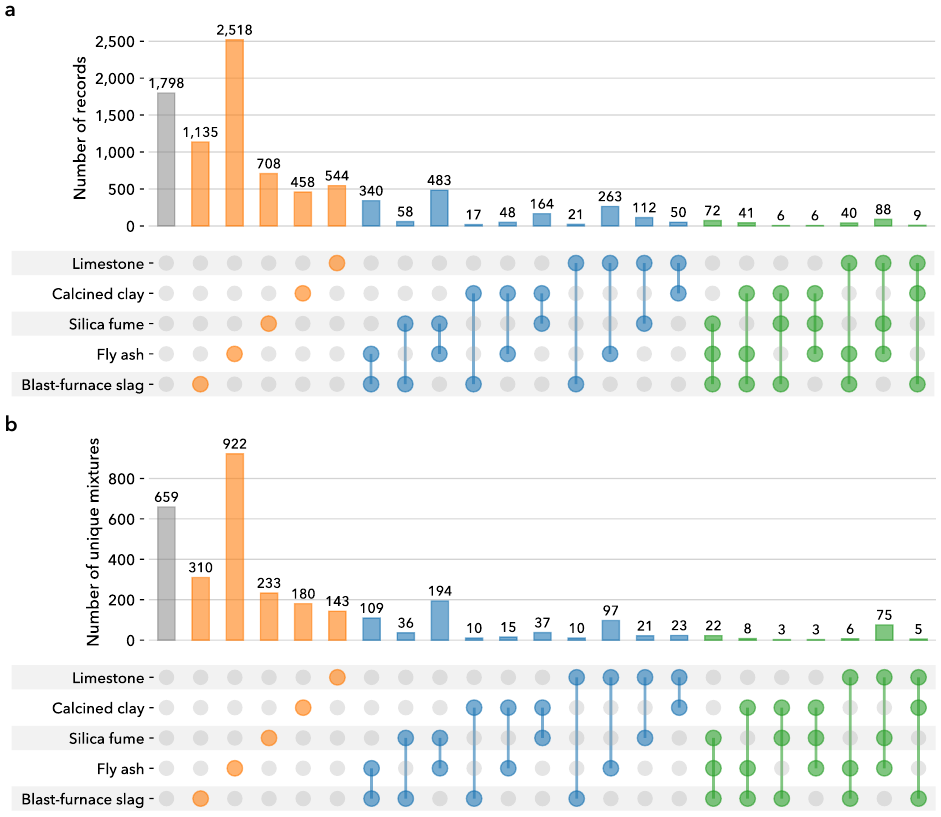}
\caption{\textbf{Distribution of records (a) and unique mixtures (b) across cementitious systems of varying complexity in the curated database.} In each subplot, filled circles denote the presence of specific supplementary cementitious materials within a system, and bars indicate the frequency corresponding to each combination.}
\label{si:fig:dataset_scm_distribution_upset}
\end{figure}

To provide a more detailed perspective, Supplementary Fig.~\ref{si:fig:dataset_scm_distribution_upset} summarizes the number of records and unique mixtures with respect to system complexity, namely, the number of SCMs incorporated in each system alongside portland cement. The results reveal that while the majority of data come from plain (no SCMs) and binary (one SCM) systems, there is also a non-negligible representation of ternary (two SCMs) and quaternary (three SCMs) systems in the curated database, which is yet to be captured in existing datasets of narrower scopes. This broader coverage of system complexity enhances the potential of the curated database to support the development of more generalizable data-driven models for diverse cementitious materials. Meanwhile, the relatively smaller number of ternary and quaternary systems underscores the persistent data scarcity in more complex cementitious systems, highlighting the need for targeted experimental and data collection efforts in future research.

As shown in Supplementary Fig.~\ref{si:fig:dataset_mixture_types_distribution}, concrete mixtures constitute the majority of both records and unique mixtures in the curated database, followed by mortar and paste. This distribution primarily reflects the current extraction scope, which focuses on tabular and textual data and on mixtures reported in mass per unit volume. Such a format is typically used for reporting concrete mixtures, whereas mortar and paste mixtures are often expressed in alternative formats (\textit{e.g.}, percentages or ratios) or presented with performance outcomes in figures rather than tables. We anticipate that the inclusion of additional data sources (\textit{e.g.}, figures) and reporting formats in future developments of the automated extraction pipeline will further enhance the representation of mortar and paste mixtures in the database.

\begin{figure}[!htb]
\centering
\includegraphics[scale=1]{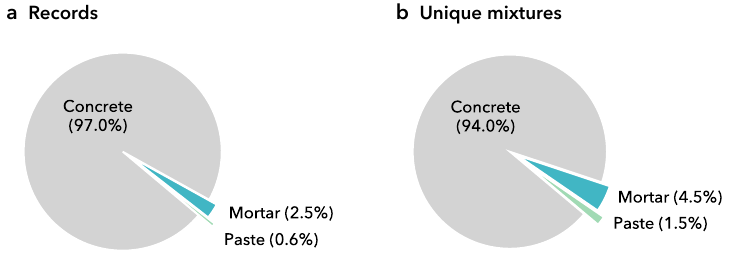}
\caption{\textbf{Distribution of records (a) and unique mixtures (b) across mixture types in the curated database.}}
\label{si:fig:dataset_mixture_types_distribution}
\end{figure}

\subsection{Environmental impact and cost analyses}

We provide here additional details on the environmental impact and cost analyses of the mixtures in the curated database, complementing the results shown in Fig.~\ref{main-fig:database}f of the main text. The analysis accounts for environmental impact and cost associated with raw material production (extraction and processing). The environmental impact of a given mixture~$i$, denoted as $E_i$, is estimated using Supplementary Eq.~(\ref{si:eq:ghg_estimation}):
\begin{equation}
E_i = \sum_{j=1}^{n} e_{j} q_{i,j} 
\label{si:eq:ghg_estimation}
\end{equation}
where $e_{j}$ is the environmental impact factor (expressed as greenhouse gas emissions in CO$_2$-equivalent per unit mass of constituent, kg\,CO$_2$-eq/kg) associated with raw material production for constituent~$j$. The term $q_{i,j}$ represents the quantity of constituent~$j$ (in kg/m$^3$) in mixture~$i$, and $n$ is the total number of mixture constituents. Similarly, the cost of a given mixture~$i$, denoted as $C_i$, is estimated using Supplementary Eq.~(\ref{si:eq:cost_estimation}):
\begin{equation}
C_i = \sum_{j=1}^{n} c_{j} q_{i,j} 
\label{si:eq:cost_estimation}
\end{equation}
where $c_{j}$ is the cost factor (expressed in USD per unit mass of constituent, USD/kg) associated with raw material production for constituent~$j$. Supplementary Table~\ref{si:tab:climate_cost_factors} summarizes the environmental impact and cost factors used in the analyses for each mixture constituent, which were obtained from the cited references following Pfeiffer \textit{et al.}\cite{Pfeiffer2024} and Nguyen \textit{et al.}\cite{Nguyen2018}. 

\begin{table}[!htb]
\centering
\caption{\textbf{Summary of environmental impact and cost factors associated with raw material production ($e_{j}$ and $c_{j}$, respectively) for each mixture constituent $j$.} Values are obtained from the cited references, following Pfeiffer \textit{et al.}\cite{Pfeiffer2024} and Nguyen \textit{et al.}\cite{Nguyen2018}.}
\label{si:tab:climate_cost_factors}
\begin{tabular}{llll}
\toprule
\textbf{$j$} & \textbf{Mixture constituent} & 
\textbf{Environmental impact} & 
\textbf{Cost factor} \\
 &  & \textbf{factor (kg\,CO$_2$-eq/kg)} & \textbf{(USD/kg)}  \\
\midrule
1 & Cement           & 1.043\cite{PortlandCementAssociation2021}   & 0.11\cite{Pfeiffer2024} \\
2 & Fly ash          & 0.00397\cite{Garcia-Segura2014} & 0.055\cite{Pfeiffer2024} \\
3 & Blast-furnace slag & 0.14705\cite{Bushi2015} & 0.11\cite{Pfeiffer2024} \\
4 & Silica fume      & 0.00397\cite{Waach2019} & 0.397\cite{Pfeiffer2024} \\
5 & Calcined clay      & 0.253\cite{Nguyen2018}  & 0.159\cite{Willett2017}  \\
6 & Limestone      & 0.037\cite{ThinkStep2015}  & 0.011\cite{U.S.GeologicalSurvey2017} \\
7 & Water            & 0.00212\cite{Wernet2016} & 0\cite{Pfeiffer2024} \\
8 & Superplasticizer            & 1.884\cite{EuropeanFederationofConcreteAdmixturesAssociationsLtd.EFCA2015} & 2.928\cite{Pfeiffer2024} \\
9 & Coarse aggregate & 0.00503\cite{Wernet2016} & 0.00992\cite{Pfeiffer2024} \\
10 & Fine aggregate   & 0.00503\cite{Wernet2016} & 0.00595\cite{Pfeiffer2024} \\
\bottomrule
\end{tabular}
\end{table}

\begin{figure}[!htb]
\centering
\includegraphics[scale=1]{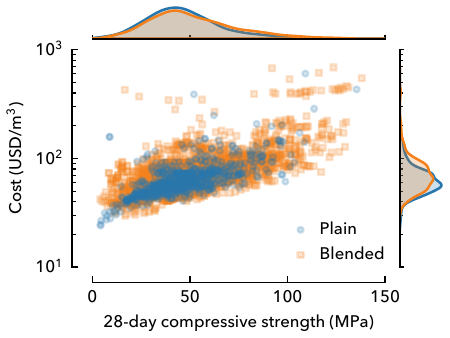}
\caption{\textbf{Cost of mixtures in the curated database as a function of 28-day compressive strength, colored by binder type.} Marginal kernel density estimates are shown along each axis to represent the data distribution.}
\label{si:fig:dataset_cost_vs_strength_bindertype}
\end{figure}

We first examine the environmental impact and cost of mixtures in the curated database (Fig.~\ref{main-fig:database}f of the main text and Supplementary Fig.~\ref{si:fig:dataset_cost_vs_strength_bindertype}, respectively). A rough positive correlation is observed between compressive strength and both environmental impact and cost, indicating that higher-performance mixtures generally incur greater environmental and economic burdens. This trend can be attributed to the increased use of cement and chemical admixtures in high-strength mixtures, which contribute significantly to both environmental impact and cost. While the blended cement mixtures generally exhibit lower environmental impact compared to plain cement mixtures at comparable strength levels, no clear trend is observed in terms of cost. This suggests that while the incorporation of SCMs effectively mitigates the environmental footprint, it does not necessarily lead to cost savings, likely due to the variations in the cost and required dosage of different SCMs. For example, blast-furnace slag was assumed to have the same cost as cement in this analysis, and silica fume is considerably more expensive (Supplementary Table~\ref{si:tab:climate_cost_factors}), which may offset the cost benefits of cement reduction in blended mixtures. The high variation of environmental impact and cost values observed for both plain and blended systems highlights substantial opportunities to optimize mixture designs for lower environmental and economic burdens without compromising performance. Conversely, suboptimal mixture proportioning may result in unnecessarily high environmental impact and cost, counteracting the original motivation for incorporating SCMs. 

We note that the environmental impact and cost factors for individual constituents may vary with region, source, technological development, market conditions, and time. More detailed analyses accounting for such variations could be undertaken in future work to assess their influence on the results. Although factors such as curing conditions and specimen dimensions also affect the measured compressive strength, they do not directly influence the environmental impact or cost; thus, the overall patterns observed here remain representative. Moreover, while the data are compiled from laboratory studies and may not fully reflect the characteristics of industrial mixtures, the observed trends and variability underscore the potential of data-driven concrete design.

\begin{figure}[!htb]
\centering
\includegraphics[scale=1]{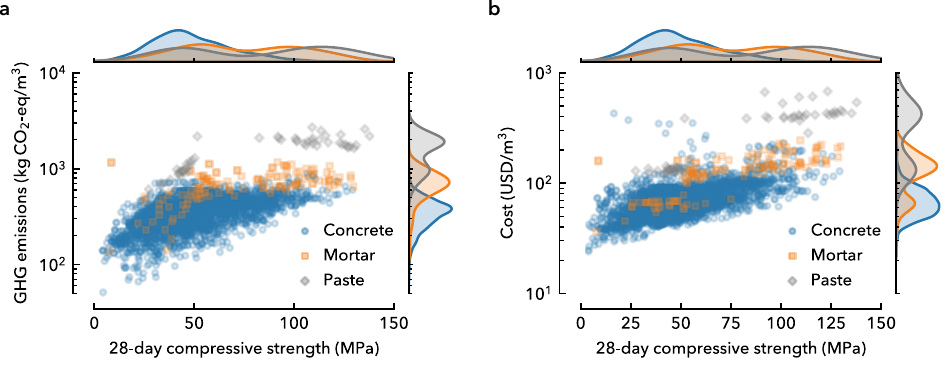}
\caption{\textbf{Environmental impact (a) and cost (b) of mixtures in the curated database as a function of 28-day compressive strength, colored by mixture type.} Marginal kernel density estimates are shown along each axis to represent the data distribution. GHG, greenhouse gas.}
\label{si:fig:dataset_co2_cost_vs_strength_mixturetype}
\end{figure}

Supplementary Fig.~\ref{si:fig:dataset_co2_cost_vs_strength_mixturetype} presents the environmental impact and cost of mixtures in the curated database across different mixture types. Both metrics generally follow the trend paste $>$ mortar $>$ concrete, reflecting the higher binder content and lower aggregate content in paste and mortar mixtures compared with concrete. The broad dispersion of environmental impact and cost values within each mixture type reinforces the importance of careful mixture proportioning and the opportunities for further optimization.

\section{Details of machine learning models} 
\label{si:sec:ml_details}

\subsection{Models}

XGBoost\cite{Chen2016a} and LightGBM\cite{Ke2017} were implemented using their respective libraries, while all other ML models, as well as the linear regression baseline, were implemented in \texttt{scikit-learn}\cite{Pedregosa2011}.

\subsection{Hyperparameter search spaces}

Supplementary Table~\ref{si:tab:ml_hyperparameter_space} summarizes the hyperparameter search space for each ML model used in this study. For all models, the number of random search iterations was fixed at 100, which was found to provide satisfactory optimization performance based on preliminary tests and prior benchmarking results reported in ref.~\cite{Grinsztajn2022}.

\begin{table}[!htb]
\centering
\caption{\textbf{Summary of hyperparameter search spaces for the machine learning models used in this study (adapted from ref.~\cite{Hollmann2025}).}}
\label{si:tab:ml_hyperparameter_space}
\begin{tabular}{lll}
\toprule
\textbf{Model} & \textbf{Hyperparameter} & \textbf{Search space} \\
\midrule
\multirow[t]{10}{*}{XGBoost\cite{Chen2016a}} 
 & \texttt{max\_depth} & $U\{1, 2, \ldots, 10\}$ \\
 & \texttt{n\_estimators} & $U\{100, 101, \ldots, 4000\}$ \\
 & \texttt{subsample} & $U(0.2, 1)$ \\
 & \texttt{colsample\_bytree} & $U(0.2, 1)$ \\
 & \texttt{colsample\_bylevel} & $U(0.2, 1)$ \\
 & \texttt{learning\_rate} & $\log U(e^{-7}, 1)$ \\
 & \texttt{min\_child\_weight} & $\log U(e^{-16}, e^{5})$ \\
 & \texttt{reg\_alpha} & $\log U(e^{-16}, e^{2})$ \\
 & \texttt{reg\_lambda} & $\log U(e^{-16}, e^{2})$ \\
 & \texttt{gamma} & $\log U(e^{-16}, e^{2})$ \\
\midrule
\multirow[t]{5}{*}{Random forest\cite{Breiman2001}}
 & \texttt{n\_estimators}        & $U\{20, 21, \ldots, 500\}$ \\
 & \texttt{max\_features}        & $\{0.3, 0.5, 0.7, \text{`log2'}, \text{`sqrt'}\}$ \\
 & \texttt{max\_depth}           & $\{\text{None}, 1, 2, \ldots, 45\}$ \\
 & \texttt{min\_samples\_split}  & $U\{2, 3, 4\}$ \\
 & \texttt{bootstrap}            & $\{\text{True}, \text{False}\}$ \\
\midrule
\multirow[t]{9}{*}{LightGBM\cite{Ke2017}}
 & \texttt{num\_leaves} & $U\{5, 6, \ldots, 50\}$ \\
 & \texttt{max\_depth} & $U\{3, 4, \ldots, 20\}$ \\
 & \texttt{n\_estimators} & $U\{50, 51, \ldots, 2000\}$ \\
 & \texttt{learning\_rate} & $\log U(e^{-3}, 1)$ \\
 & \texttt{subsample} & $U(0.2, 0.8)$ \\
 & \texttt{colsample\_bytree} & $U(0.2, 0.8)$ \\
 & \texttt{min\_child\_weight} & $\{10^{-5}, 10^{-3}, 10^{-2}, 10^{-1}, 1, 10, 10^{2}, 10^{3}, 10^{4}\}$ \\
 & \texttt{reg\_alpha} & $\{0, 10^{-1}, 1, 2, 5, 7, 10, 50, 100\}$ \\
 & \texttt{reg\_lambda} & $\{0, 10^{-1}, 1, 5, 10, 20, 50, 100\}$ \\
\midrule
\multirow[t]{7}{*}{Multilayer perceptron\cite{Rumelhart1986}}
 & \texttt{hidden\_layer\_sizes} & Depth $\in\{1,2,3\}$; \\
 &  &  each layer width sampled from $U\{16,\ldots,264\}$ \\
 & \texttt{activation} & $\{\text{`relu'}, \text{`tanh'}\}$ \\
 & \texttt{alpha} & $\log U(e^{-7}, 0.1)$ \\
 & \texttt{learning\_rate\_init} & $\log U(e^{-4}, 0.5)$ \\
 & \texttt{early\_stopping} & $\{\text{True}, \text{False}\}$ \\
 & \texttt{max\_iter} & $\{5000\}$ \\
\midrule
\multirow[t]{3}{*}{Support vector machine\cite{Cortes1995}}
 & \texttt{C} & $\{0.1, 1, 10, 100\}$ \\
 & \texttt{gamma} & $\{\text{`auto'}, \text{`scale'}\}$ \\
 & \texttt{kernel} & $\{\text{`rbf'}, \text{`poly'}, \text{`sigmoid'}\}$ \\
\bottomrule
\end{tabular}
\end{table}

\subsection{Feature importance analysis}

\begin{figure}[!htb]
\centering
\includegraphics[scale=1]{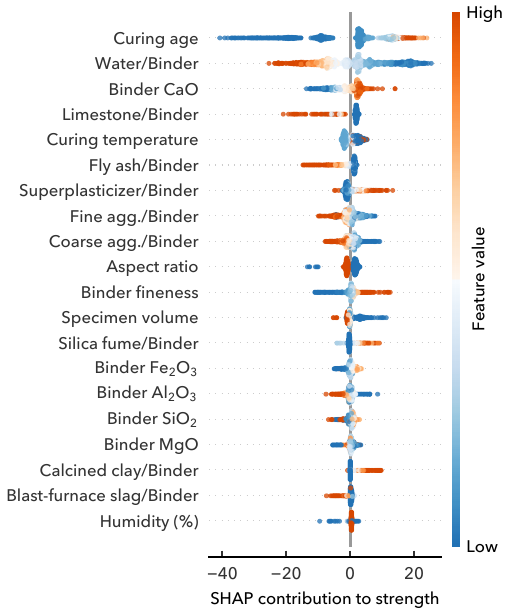}
\caption{\textbf{Summary plot of feature importance, showing the distribution of SHapley Additive exPlanations (SHAP) values for all input features of the XGBoost model with additional chemical and physical descriptors.} The plot for the 12 most important features is shown in Fig.~\ref{main-fig:ml}b of the main text.}
\label{si:fig:ml_shap_full_xgb}
\end{figure}

Supplementary Fig.~\ref{si:fig:ml_shap_full_xgb} presents the complete summary plot of feature importance based on SHapley Additive exPlanations (SHAP) values for all input features of the XGBoost model with additional chemical and physical descriptors, complementing the summary of the most influential features shown in Fig.~\ref{main-fig:ml}b of the main text. Overall, the relationships between feature values and their corresponding SHAP values are generally consistent with established domain knowledge, indicating that the model captures physically meaningful trends. Beyond the top features highlighted in the main text, other features related to chemical composition also contribute meaningfully to the model predictions, albeit to a lesser extent. These observations further underscore the value of incorporating detailed chemical and physical descriptors in ML modeling, and motivate future studies to more systematically investigate their roles in governing mixture performance and to develop more informative descriptors that better capture the underlying material behavior.

\subsection{Additional evaluation results}

\begin{figure}[!htb]
\centering
\includegraphics[scale=1]{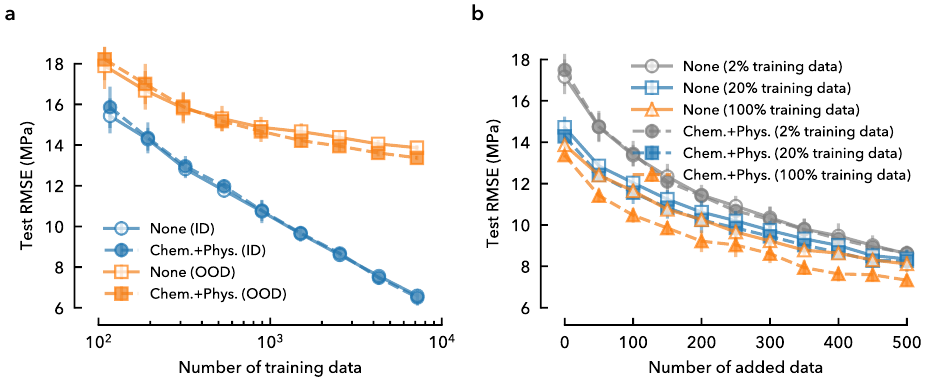}
\caption{\textbf{In-distribution (ID) and out-of-distribution (OOD) performance of the random forest model.} \textbf{a}, ID and OOD performance as a function of training data size.
\textbf{b}, OOD performance when varying amounts of test-domain data are added to the training set. The OOD task definition and evaluation protocol are identical to those described in Fig.~\ref{main-fig:ml}c, d of the main text. Results are averaged over 10 runs with different random seeds, with error bars indicating standard deviation.}
\label{si:fig:ml_performance_rf}
\end{figure}

\begin{figure}[!htb]
\centering
\includegraphics[scale=1]{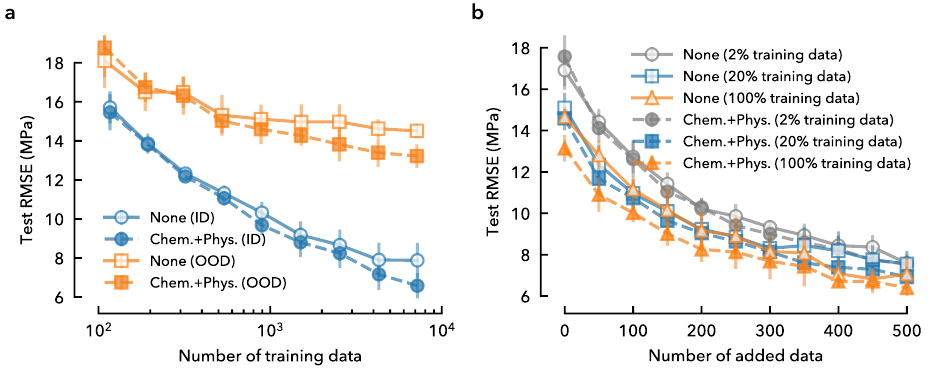}
\caption{\textbf{In-distribution (ID) and out-of-distribution (OOD) performance of the LightGBM model.} \textbf{a}, ID and OOD performance as a function of training data size.
\textbf{b}, OOD performance when varying amounts of test-domain data are added to the training set. The OOD task definition and evaluation protocol are identical to those described in Fig.~\ref{main-fig:ml}c, d of the main text. Results are averaged over 10 runs with different random seeds, with error bars indicating standard deviation.}
\label{si:fig:ml_performance_lgbm}
\end{figure}

Supplementary Figs.~\ref{si:fig:ml_performance_rf} and \ref{si:fig:ml_performance_lgbm} present the in-distribution (ID) and out-of-distribution (OOD) performance of two additional leading ML models---random forest and LightGBM, respectively---complementing the XGBoost results shown in Fig.~\ref{main-fig:ml}c, d of the main text. Despite modest differences in absolute performance across models, the overall trends in ID and OOD performance as a function of training data size, as well as the OOD performance gains with the addition of test-domain data, are consistent across all three models. These results further reinforce the importance of large and diverse training datasets, as well as the inclusion of informative input features, for achieving robust and generalizable ML models in concrete materials design.

\backmatter